\documentclass[a4paper,12pt]{article}\usepackage[]{graphicx}\usepackage[]{color}
\makeatletter
\def\maxwidth{ %
  \ifdim\Gin@nat@width>\linewidth
    \linewidth
  \else
    \Gin@nat@width
  \fi
}
\makeatother

\definecolor{fgcolor}{rgb}{0.345, 0.345, 0.345}
\newcommand{\hlnum}[1]{\textcolor[rgb]{0.686,0.059,0.569}{#1}}%
\newcommand{\hlstr}[1]{\textcolor[rgb]{0.192,0.494,0.8}{#1}}%
\newcommand{\hlcom}[1]{\textcolor[rgb]{0.678,0.584,0.686}{\textit{#1}}}%
\newcommand{\hlopt}[1]{\textcolor[rgb]{0,0,0}{#1}}%
\newcommand{\hlstd}[1]{\textcolor[rgb]{0.345,0.345,0.345}{#1}}%
\newcommand{\hlkwa}[1]{\textcolor[rgb]{0.161,0.373,0.58}{\textbf{#1}}}%
\newcommand{\hlkwb}[1]{\textcolor[rgb]{0.69,0.353,0.396}{#1}}%
\newcommand{\hlkwc}[1]{\textcolor[rgb]{0.333,0.667,0.333}{#1}}%
\newcommand{\hlkwd}[1]{\textcolor[rgb]{0.737,0.353,0.396}{\textbf{#1}}}%

\usepackage{framed}
\makeatletter
\newenvironment{kframe}{%
 \def\at@end@of@kframe{}%
 \ifinner\ifhmode%
  \def\at@end@of@kframe{\end{minipage}}%
  \begin{minipage}{\columnwidth}%
 \fi\fi%
 \def\FrameCommand##1{\hskip\@totalleftmargin \hskip-\fboxsep
 \colorbox{shadecolor}{##1}\hskip-\fboxsep
     \hskip-\linewidth \hskip-\@totalleftmargin \hskip\columnwidth}%
 \MakeFramed {\advance\hsize-\width
   \@totalleftmargin\z@ \linewidth\hsize
   \@setminipage}}%
 {\par\unskip\endMakeFramed%
 \at@end@of@kframe}
\makeatother

\definecolor{shadecolor}{rgb}{.97, .97, .97}
\definecolor{messagecolor}{rgb}{0, 0, 0}
\definecolor{warningcolor}{rgb}{1, 0, 1}
\definecolor{errorcolor}{rgb}{1, 0, 0}
\newenvironment{knitrout}{}{} 

\usepackage{alltt}
\usepackage[utf8x]{inputenc}
\usepackage{natbib}
\usepackage{amsmath}
\usepackage{amssymb}
\usepackage{amsthm}
\usepackage{hyperref}
\usepackage{graphicx}
\usepackage{color}
\usepackage{multirow}
\usepackage{booktabs}
\usepackage[]{threeparttable}
\usepackage[left = 3cm, right = 2.5cm, bottom = 2cm, top = 3cm]{geometry}
\usepackage{tikz}

\def\tr{{\rm tr}}
\def\pr{{\rm Pr}}

\def\diag{{\rm diag}}

\newtheoremstyle{example}
{3pt} 
{3pt} 
{} 
{0\parindent} 
{\bf}
{:} 
{.5em} 
{} 
\newtheoremstyle{theorem}
{3pt} 
{3pt} 
{\em} 
{0\parindent} 
{\bf}
{:} 
{.5em} 
{} 
\theoremstyle{example} 
\theoremstyle{theorem}

\title{Accurate inference in  negative binomial regression}
\author{E. C. KENNE PAGUI, A. SALVAN and N. SARTORI \\
\small University of Padova, Department of Statistical Sciences\\
\small kenne@stat.unipd.it, salvan@stat.unipd.it,  sartori@stat.unipd.it
}
\date{}  

\IfFileExists{upquote.sty}{\usepackage{upquote}}{}
\begin{document}

\maketitle

\begin{abstract} 
\noindent
Negative binomial regression is commonly employed to analyze overdispersed count data. With small to moderate sample sizes, the maximum likelihood estimator of the dispersion parameter may be subject to a significant bias, that in turn affects inference on mean parameters.  This paper proposes inference for negative binomial regression based on adjustments of the score function aimed at  mean and median bias reduction. 
The resulting estimating equations are similar to those available for improved inference in generalized linear models and, in particular, can be solved using a suitable extension of iterative weighted least squares. Simulation studies show a remarkable performance of the new methods, which are also found to solve in many cases numerical problems of maximum likelihood estimates. The methods are illustrated and evaluated using two case studies: an  Ames salmonella assay data set and data on epileptic seizures. Inference based on adjusted scores turns out to be generally preferable to explicit bias correction.

  \noindent 
\end{abstract}

\noindent
\emph{Some key words:} adjusted score; iterative weighted least squares; maximum likelihood; mean and median bias reduction; parameterization invariance.

\section{Introduction}
\label{intro}
Regression models for count data  are very common in many contexts, especially in 
social sciences,  economics, biology and epidemiology. It is not uncommon that empirical counts display substantial overdispersion and a popular modelling approach is negative binomial regression, see e.g.\ \citet[][Section 7.3]{agre2015} and \citet{hilbe2011} for recent accounts. 


Frequentist inference about mean and dispersion parameters in negative binomial regression is tipically based on the likelihood and this is the method of choice for standard software, such as the {\tt glm.nb} function of the  R package MASS \citep{venables2002}. 
Maximum likelihood  has been studied starting from \citet{fisher1941} and \citet{ansco1950} for independent and identically distributed data  and from \citet{lawl1987} for the regression setting. In particular, existence conditions for the maximum likelihood estimate, extending the random sampling condition that the empirical variance has to exceed the sample mean, are given in  \citet{gning2013}.


With moderate sample sizes, the maximum likelihood estimator of the dispersion parameter may be subject to a substantial bias that can influence the inferential conclusions. 
For independent and identically distributed data, \citet{saha2005b} derive a bias corrected maximum likelihood  estimator  and show that it is preferable to other methods considered in \citet{clark1989} and in \citet{piego1990}. 
The Authors also give an example involving negative binomial regression, but do not provide the expression of the estimator.

For generalized linear models, improvement to maximum likelihood can be achieved both by explicit mean bias correction and by adjusting the likelihood equations, resulting in mean or median bias reduction \citep{kosmietal2017}. Bias reduction was introduced by \citet{firth1993} \citep[see also][]{kosmi2009, kosmi2010}.   Median bias reduction, developed by 
\citet{kenne2017}, is such that each component of the estimator is, with high accuracy, median unbiased, that is, it has the same probability of underestimating and overestimating the corresponding parameter component. 
Mean and median bias reduced estimators  have smaller mean and median bias than the  maximum likelihood estimator, respectively. Mean bias reduction is invariant under linear transformation of the parameters, while median bias reduction  is invariant under monotone component-wise transformations of the parameters.  

In this paper, we extend to negative binomial regression the improved maximum likelihood methods of \citet{kosmietal2017} and of \citet{saha2005b}.  As in \citet{kosmietal2017} for  generalized linear models, we derive explicit formulae for the adjusted score equations and show that they can be solved by iterative weighted least squares after an  appropriate adjustment of the ordinary working variates for maximum likelihood. Moreover, the method is developed for various link functions and  parameterizations of the dispersion parameter. An \texttt{R}  implementation is given in the  \texttt{brnb} function available  in the forked  \texttt{brglm2}   \texttt{R}  package \citep{brglm2} on GitHub (\url{https://github.com/eulogepagui/brglm2}). 

All the proposed estimators are compared  through an extensive simulation experiment  under different scenarios and through two case studies, the Ames salmonella reverse mutagenicity assay  presented in \citet{marg1989} and the epileptic seizures data from  \citet{thall1990}.  The results indicate that  mean and median bias reduction both outperform standard likelihood inference, especially with moderate sample sizes. Median bias reduction  provides the best empirical coverage of Wald-type confidence intervals for all parameters.
Moreover, numerical problems that lead to unavailability of the maximum likelihood estimate, and therefore of its bias correction,  occur more frequently than with mean or median bias reduction. 
In addition, bias correction is seen to be less accurate than implicit methods when the number of covariates is large relative to the sample size.



The rest of the paper is organized as follows. In Section 2, we introduce the notation for the negative binomial regression model.  In  Section 3 we give the adjusted score functions for mean and median bias reduction, together with computational details. Sections 4 and 5 contain simulation results and case studies, respectively.  The Supplementary Material contains additional figures and the \texttt{R} code to reproduce the analyses in the paper.

\section{Negative binomial regression}
\label{sec:model}
 Let $y_i$, $i=1,\ldots,n$, be  realizations of independent negative binomial random variables $Y_i$ with mean
$\mu_i$, variance $V(Y_i)=\mu_i+\kappa\mu_i^2$, where $\kappa >0$ is a dispersion parameter. The probability mass function is   
\begin{equation}\label{model}
f_{Y_i}(y_i;\mu_i,\kappa)=\frac{\Gamma(y_i+\kappa^{-1})}{y_i!\Gamma(\kappa^{-1})} \left( \frac{\kappa\mu_i}{1+\kappa\mu_i} \right)^{y_i}
 \left( \frac{1}{1+\kappa\mu_i} \right)^{1/\kappa}\,,
 \end{equation}
$y_i=0,1,\ldots$, $\kappa>0$ and $\mu_i>0$.

In a regression setting, we consider $\mu_i=g^{-1}(\eta_i)$, where $g^{-1}(\cdot)$ is the inverse of the link function, $\eta_i= x_i \beta$ is the linear predictor, with $\beta=(\beta_1,\cdots,\beta_p)^\top \in\mathbb{R}^p$ and $x_i=(x_{i1},\ldots, x_{ip})$  a row vector of covariates. The usual choice for the link function is $g(\cdot) = \log(\cdot)$.
For sake of generality, the derivation below is for a generic smooth reparameterization of $\kappa$, say $\phi=\phi(\kappa)$ with inverse $\kappa(\phi)$ and derivative with constant sign $\kappa'(\phi)$. Common choices are $\phi=1/\kappa$,  
$\phi=\log\kappa$ and $\phi=\sqrt{k}$. 

 Let  $\theta=(\beta^\top,\phi)^\top$. 
Noting that  for any $a >0$, $\Gamma(y+a)/\Gamma(a)=a(a+1)\cdots(a+y-1)$, the log likelihood is 
$$
\ell(\beta,\phi)= \sum_{i=1}^n m_i\left\{ \sum_{j=0}^{y_i^*}\log(1+\kappa j) +y_i\log\frac{\mu_i}{1+\kappa\mu_i}-\frac{1}{\kappa}\log(1+\kappa\mu_i) \right\},
$$
where $m_i$ is a fixed prior weight for the $i$th observation, $y_i^*=y_i-1$, $ \sum_{j=0}^{y_i^*}$ is zero when $y_i^*<0$ and $\kappa=\kappa(\phi)$. 

The score function  $U=U(\theta)=(\partial/\partial \theta) \ell(\theta)$ has components $U_\beta = (\partial/\partial\beta)$ 
$\ell(\beta,\kappa(\phi))$ and $U_\phi= (\partial/\partial\phi) \ell(\beta,\kappa(\phi))$ given by 
\begin{align*}
&U_{\beta}=X^\top WD^{-1}(y-\mu),\\
&U_{\phi}=  \kappa'(\phi)\sum_{i=1}^n m_i \left\{  S_{1i} -\frac{\mu_i y_i}{\kappa\mu_i+1} + \frac{(\kappa\mu_i+1)\log(\kappa\mu_i+1)-\kappa\mu_i}{\kappa^3\mu_i+\kappa^2}\right\},
\end{align*}
where $D$ is a diagonal matrix with generic entry $d_{i}=d\mu_i/d\eta_i$, 
$W$ is a diagonal matrix with generic entry $w_i=m_id_{i}^2/V(Y_i)$ (the $i$th working weight), $y=(y_1,\cdots,y_n)^\top$, $\mu=(\mu_1,\cdots,\mu_n)^\top$ and 
$S_{1i}= \sum_{j=0}^{y_i^*} j/(\kappa j+1)$. 

 The expected information,  obtained in \citet{lawl1987}, is
 \begin{align*}
& i(\theta) =
\left[
\begin{array}{cc}
i_{\beta\beta} & 0_p \\
0_p^\top & i_{\phi\phi}
\end{array}
\right]
=
\left[
\begin{array}{cc}
X^\top W X & 0_p \\
0_p^\top &\kappa'(\phi)^2 i_{\kappa\kappa} \end{array}
\right]\,,
 \end{align*}
where $0_p$ is a $p$-vector of zeros and 
$$
i_{\kappa\kappa}= \kappa^{-4} \sum_{i = 1}^n m_i \left\{ \sum_{j=0}^{+\infty} 
\frac{\pr\left(Y_i > j\right) }{(\kappa^{-1}+j)^2}  -\frac{\kappa \mu_i}{\mu_i+\kappa^{-1}}\right\}\,.
$$


The maximum likelihood estimate $\hat\theta^\top=(\hat\beta^\top, \hat\phi)$ is obtained as solution of the equations  $U_{\beta}=0$ and $U_{\phi}=0$ that can be solved using a Fisher scoring algorithm. 
Exploiting the orthogonality between $\beta$ and $\phi$, the current iterate $\hat\phi^{(j)}$ is found by replacing  $\hat\beta^{(j)}$  into the $j$th  Fisher scoring iteration for  $U_{\phi}=0$. The procedure is alternated until convergence.
 With simple algebra,
the $j$th iteration of  Fisher scoring algorithm  for $U_{\beta}=0$ 
updates the current iterate $\hat\beta^{(j)}$ providing
\begin{equation}\label{2}
\hat\beta^{(j+1)}  =  (X^\top W^{(j)}X)^{-1}X^\top W^{(j)} z^{(j)},
\end{equation}
where the superscript $(j)$ indicates that the quantity is evaluated at $\hat\beta^{(j)}$ and the generic element $z_i=\eta_i+(y_i-\mu_i)/d_{i},\, i=1,\ldots,n$, of $z$ is usually called the adjusted dependent variable or working variate.  This has the same form of the iterative weighted least squares (IWLS) iteration used in generalized linear models.

\section{Mean and median bias reduction}
\label{sec:br_inference}

Bias of maximum likelihood estimators in small samples or with sparse data can result in significant loss of accuracy of the related inferential procedures. An extensive amount of literature has focused on methods for reducing such bias either explicitly, that is subtracting from the maximum likelihood estimate an estimate of its  first order bias, or implicitely, that is modifying the estimating function.  We refer to \citet{kosmidis2014} for a  unified review. See also \citet{greenlandetal2016} for an expository discussion of sparse data bias and available remedies. 

We recall below the various analytic improvements to maximum likelihood and 
obtain their expression for negative binomial regression. 

For a regular model with $d$-dimensional parameter $\theta$,   log likelihood $\ell(\theta)$,   score function $U(\theta)$,  the expected information  $i(\theta)$ is the   assumed  to be of order $n$. We let  $U_{\theta_r}(\theta)$ be a generic component of $U(\theta)$, $r=1,\ldots,d$, and  $j(\theta)=-\partial^2\ell(\theta)/\partial\theta\partial\theta^\top$ be the observed information. 

All the proposed adjustments involve the quantities
$$
P_{\theta_r}(\theta)=E_{\theta}\{U(\theta)U(\theta)^\top U_{\theta_r}(\theta)\},\;\;
Q_{\theta_r}(\theta)=E_{\theta}\{-j(\theta) U_{\theta_r}(\theta)\},\;\; r=1,\ldots,d.
$$ 
Score adjustments have the general form 
$U(\theta) + A(\theta)$, 
with $A(\theta)$ a model-dependent modification term of order $O(1)$ under repeated sampling. The modified maximum likelihood estimator is obtained as solution of $U(\theta) + A(\theta)=0$ and, being the correction of order $O(1)$, it has the same asymptotic distribution as the maximum likelihood estimator. In practice, standard errors are computed using diagonal elements of the inverse of the expected  information, evaluated at the modified estimate.

In particular, bias of order $O(n^{-1})$ of the maximum likelihood estimator $\hat\theta$ is implicitly  removed \citep{firth1993} with 
$A(\theta) = A^*(\theta)$,  where $A^*(\theta)$  has generic entry
\begin{equation}\label{A*}
A^*_{\theta_r}(\theta) = \frac{1}{2} \tr\left\{i(\theta)^{-1} (P_{\theta_r}(\theta)+Q_{\theta_r}(\theta))\right\}\,.
\end{equation}
We let $U^*(\theta)=U(\theta) + A^*(\theta)$ and we denote by $\theta^*$ the corresponding estimator, solution of $U^*(\theta)=0$.

The  explicitly bias corrected maximum likelihood estimate,  see e.g.\ \citet[][Section 9.2(vii)]{cox:hink:1974}  and \citet[][Section 5.3]{barn:cox:1994} is given by 
$
\tilde\theta=\hat\theta-b(\hat\theta)$, 
where $b(\theta) =-i(\theta)^{-1}A^*(\theta)$ is the $O(n^{-1})$ bias of $\hat\theta$. 

Both $\theta^*$ and $\tilde\theta$ have bias of order $O(n^{-2})$. 
When $\theta$ is the canonical parameter
of a full exponential family, $\theta^*$  is the mode of the posterior distribution obtained  using Jeffreys' prior. 
However, both bias reduction and bias correction are tied to a specific parameterization. This means that if $\psi=\psi(\theta)$ is a nonlinear reparameterization of $\theta$, the transformed estimator $\psi(\theta^*)$ or $\psi(\tilde\theta)$ will not have reduced bias of order $O(n^{-2})$.

Equivariance under nonlinear componentwise reparameterizations is obtained with median bias reduction \citep{kenne2017}, leading to the estimator $\theta^\dagger$ satisfying, in the continuous case,  the improved median centering property 
$
\pr_\theta(   \theta_r^\dagger \leq \theta_r )=1/2 + O(n^{-3/2})$,\, $r=1,\ldots,d$,
in contrast with the corresponding $O(n^{-1/2})$ order of error for the maximum likelihood estimator. Median bias reduction is achieved using
$A(\theta) = A^\dagger(\theta)$,  given in \citet[][formula (10)]{kenne2017}.

For the negative binomial regression model  (\ref{model}), we have $d=p+1$ and the quantity $A^*(\theta)$, whose derivation is in the Appendix, has blocks
 $$
 A_\beta^*=X^\top W\xi\,,\qquad  A_\phi^*=\kappa'(\phi)\sum_{i=1}^n \frac{m_ih_id_{i}^2\mu_i^2}{2w_iV(Y_i)^2}+\frac{1}{2}i_{\phi\phi}^{-1}R_{\phi\phi}\,,
 $$
where  $\xi =(\xi_1,\ldots,\xi_n )^\top$, with $\xi_i = h_id_i'/(2d_iw_i)$. 
The quantity $h_i$ appearing in $\xi_i$ and in $A_\phi^*$  is
the `hat' value for the $i$th observation, obtained as the $i$th diagonal element of the matrix $H = X (X^\top W X)^{-1} X^\top W$  and  $d_i'=d^2\mu_i/d\eta_i^2$. The expression of $R_{\phi\phi}$ is given in the Appendix. 

The median bias adjustment $A^\dagger(\theta)$ for negative binomial regression has blocks
 \begin{equation}\label{adjglm}
 A_{\beta}^\dagger =  X^\top W (\xi + X u)\,, \qquad  \quad A_\phi^\dagger = A^*_\phi+i_{\phi\phi}^{-1}S_{\phi\phi} \, ,
\end{equation}
 where expressions for $u$ and  $S_{\phi\phi}$ are given in the Appendix. 

With simple algebra,  the $j$th iteration of   IWLS which updates the current iterate $\beta^{*(j)}$ leads to 
\begin{equation}\label{5}
\beta^{*(j+1)}  =  (X^\top W^{(j)}X)^{-1}X^\top W^{(j)} z^{*(j)}\,,
\end{equation}
where $z^{*(j)}= z^{(j)}+\xi^{(j)}$ is the adjusted version of the working variate $z$ defined in (\ref{2}). 
The $j$th iteration of  IWLS for $\tilde\theta$ has the same expression as (\ref{5}), with
adjusted version of working variate  $\tilde z= z^*+ Xu$.


 All the improved methods for negative binomial regression, together with maximum likelihood fitting, are available in the {\tt brnb} R function  in the forked  \texttt{brglm2}   \texttt{R}  package \citep{brglm2} on GitHub (\url{https://github.com/eulogepagui/brglm2}). 
Maximum likelihood fitting can also be performed for instance using the {\tt glm.nb} function of the MASS  \texttt{R} library.

\section{ Simulation studies}
\label{sec:simulations}


In this section, the properties of the estimators are assessed through simulation under different  scenarios corresponding to combination of values of   $n$, $\phi$, and $\beta$. For each scenario, we run 10000 Monte Carlo replications. In all cases, we use the logarithmic link function and  the identity transformation for the dispersion parameter $(\phi=\kappa)$. We compute estimates using maximum likelihood (ML), mean and median bias reduction  (BR) through the \texttt{brnb} \texttt{R} function. 

Estimators are evaluated in terms of
empirical probability of underestimation (PU), estimated relative (mean) bias (RBIAS), estimated coverage probability of 95\% Wald-type confidence intervals (WALD) and the relative increase in mean squared error from its absolute minimum due to bias (IBMSE) given by $100 B^2/SD^2$. Here, $B$ denotes the estimated mean bias and $SD$, the corresponding estimated standard deviation. The four performance measures are expressed in percentages. 

\begin{figure}[!ht]
\centering
\caption{Estimated relative bias (RBIAS), probability of under estimation (PU) and  coverage probability of 95\% Wald-type confidence intervals (WALD) for the intercept $\beta=\log\mu$. Results for ML (black squares), mean BR (blue circles) and median BR (red triangles).}\label{fig1}
\vspace{0.2cm}
\resizebox{\textwidth}{!}{
\begin{tabular}{c}
\includegraphics[]{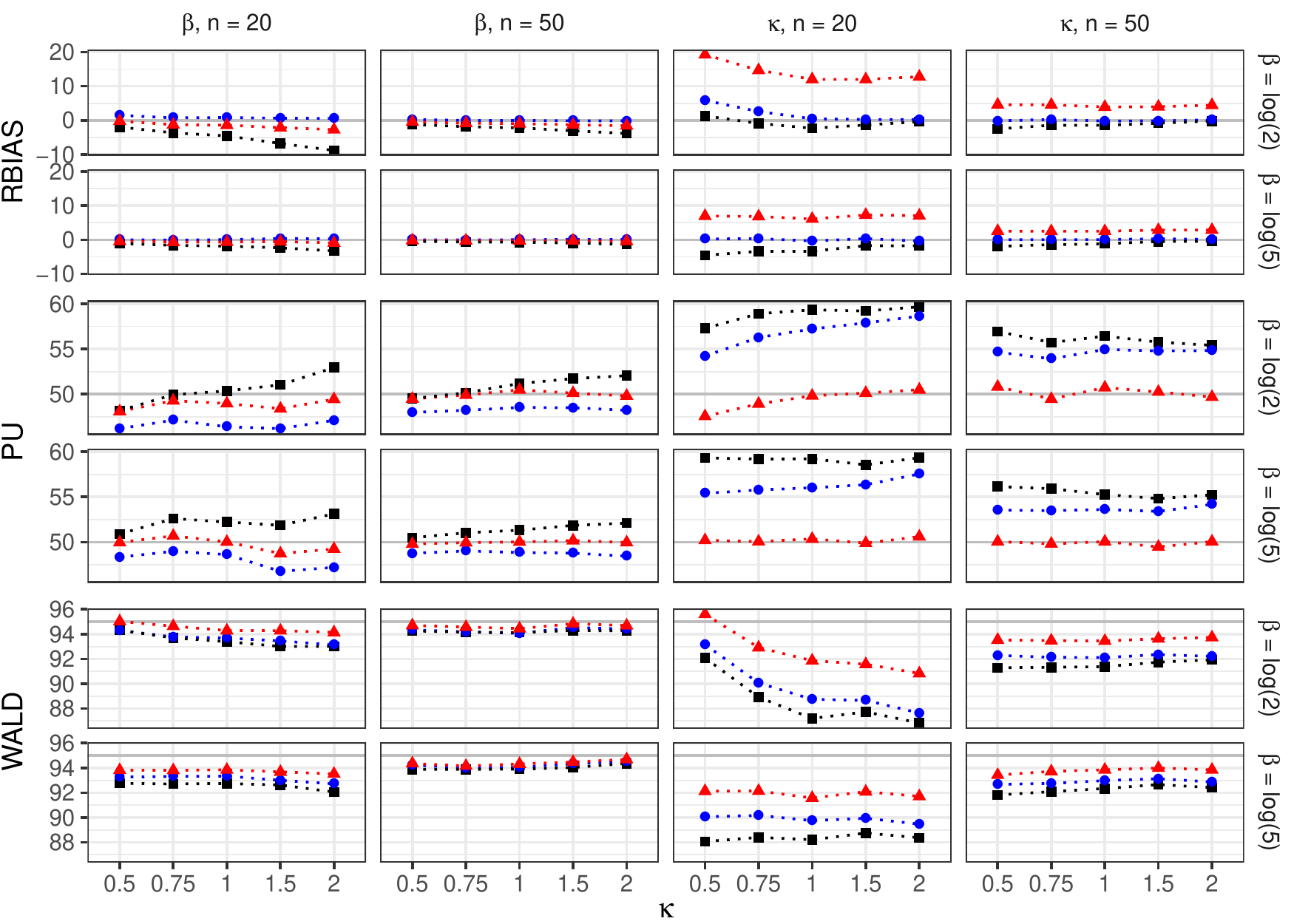}
\end{tabular}
}
\end{figure}

We first  conduct a simulation study with constant mean $\mu$, i.e.\ with intercept only.   Mean bias reduction with a numerical example for this case was considered  in \citet[][Example 4]{zhang2019}. In particular, we take sample sizes $n=20$ and $n=50$ from $NB(\mu,\kappa)$ for the combinations of $\mu=2,5$ $(\beta=\log 2,\log 5)$ and $\kappa=0.5,0.75,1,1.5,2$.
\begin{table}[!ht]
\centering
\caption{Computational diagnostics in 10000 replications. $A_1$ indicates the number of samples with empirical variance less than mean. Of the remaing samples,  $A_2$ is the number of non  convergence samples using ML, $A_3$ is the number of non  convergence samples using mean BR and  $A_4$ is the number of non  convergence samples using median BR.}\label{ conv }
\vspace{0.1cm}
\begin{tabular}{| l | l | ccccc | ccccc |}
  \cline{3-12}
 \multicolumn{2}{}{}& \multicolumn{5}{|c|}{$\mu=2$} & \multicolumn{5}{c|}{$\mu=5$}\\
\cline{2-12}
 \multicolumn{1}{c|}{}& $\kappa$ & 0.5 & 0.75 & 1 & 1.5 & 2 & 0.5 & 0.75 & 1 & 1.5 & 2 \\ 
  \hline
\multirow{4}{*}{$n=20$}& $A_1$ & 535 & 214 & 108 & 43 & 17 & 18 & 3 & 0 & 0 & 0 \\ 
  &$A_2$ & 163 & 85 & 42 & 17 & 12 & 6 & 3 & 0 & 3 & 1 \\ 
  &$A_3$ & 2 & 1 & 0 & 0 & 1 & 0 & 0 & 0 & 2 & 1 \\ 
  &$A_4$ & 0 & 0 & 0 & 0 & 0 & 0 & 0 & 0 & 2 & 1 \\ 
  \hline
 \multirow{4}{*}{$n=50$}&  $A_1$ & 36 & 6 & 0 & 0 & 0 & 0 & 0 & 0 & 0 & 0 \\ 
  &$A_2$ & 16 & 0 & 1 & 0 & 0 & 0 & 0 & 0 & 0 & 0 \\ 
  &$A_3$ & 2 & 0 & 0 & 0 & 0 & 0 & 0 & 0 & 0 & 0 \\ 
  &$A_4$ & 0 & 0 & 0 & 0 & 0 & 0 & 0 & 0 & 0 & 0 \\ 
   \hline
\end{tabular}
\end{table}
For each setting, Table \ref{ conv } gives the number of samples, out of 10000 replications, with variance less than the mean and occurrence of non convergence. The results are presented in Figure \ref{fig1}. For each method the results are reported only for samples with convergence. Therefore, the performance for maximum likelihood should be judged with caution.
Both
mean and median BR achieve the desired goals, i.e.\ are effective in mean and median centering, respectively,  and are both preferable to ML. All three estimators improve as the sample size and $\mu$ increase. In particular, median BR  provides empirical coverage of the 95\% Wald-type confidence intervals  closest to nominal. We finally note that ML for the dispersion parameter has smaller mean bias than median BR in this parameterization, but not when  the  inverse or the log parameterization  is adopted. On the other hand, median BR is not targeted for reduction of the mean bias. 

%
%

We now  consider a second simulation study involving covariates.  In particular, we let 
\begin{equation}
\log\mu_i = \beta_1+\beta_2 x_{i2} +\beta_3 x_{i3}+\beta_4 x_{i4}+\beta_5 x_{i5},
\end{equation}
where $x_{i2}$ are  $n$ independent realizations of Bernoulli $B(1,0.8)$; $x_{i3}$ is generated from a Bernoulli $B(1,0.5)$; $x_{i4}$ is generated  from a uniform $U(1,2)$;   $x_{i5}$ is generated  from a Poisson $P(2.5)$, $i=1,\ldots,n$. The true parameter values are $\beta_1=1,\beta_2=-0.75,\beta_3=-1.5,\beta_4=1$ and $\beta_5=-0.5$. Four values were considered for the  dispersion parameter, $\kappa=0.5,0.75,1,1,5$.
The sample sizes considered were $n = 40, 80$. For each combination of $\beta,\kappa$ and $n$, we
run 10000 Monte Carlo replications, where the values of the explanatory variables $x_{i2}, x_{i3}, x_{i4}$ and $x_{i5}$   were held constant throughout the simulations.


\begin{figure}[!ht]
\centering
\caption{Estimation of regression parameters $\beta=(\beta_1,\beta_2,\beta_3,\beta_4,\beta_5)$ with $\kappa=0.75,n=40,80$. Simulation results for ML (black squares), mean BR (blue circles) and median BR (red triangles). }\label{fig2}
\vspace{0.2cm}
\resizebox{\textwidth}{!}{
\begin{tabular}{c}
\includegraphics[]{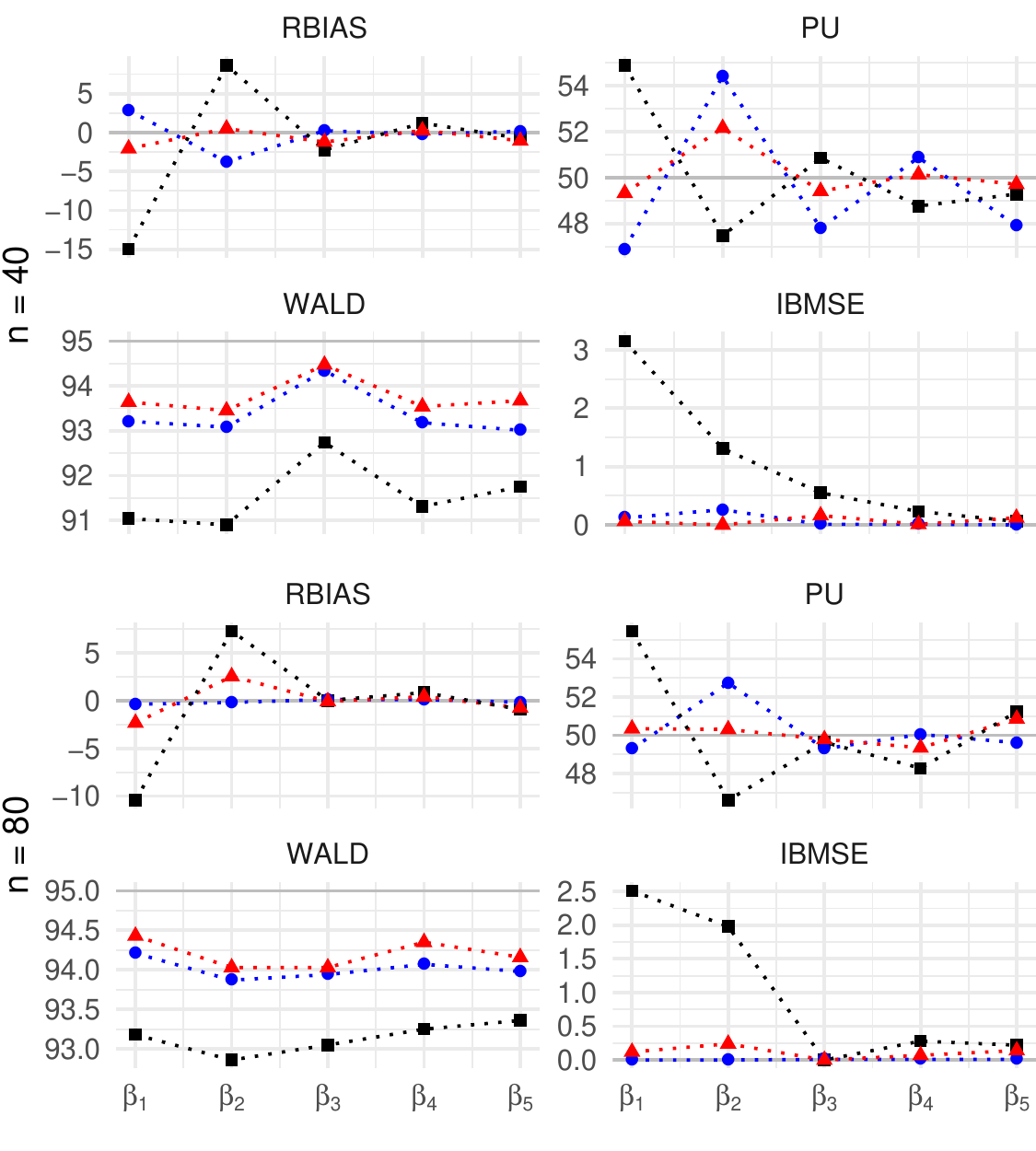}
\end{tabular}
}
\end{figure}

%
%
%
\begin{figure}[!ht]
\centering
\caption{Estimation of dispersion parameter $\kappa$ with $n=40,80$. Simulation results for ML (black squares), mean BR (blue circles) and median BR (red triangles). }\label{fig3}
\vspace{0.2cm}
\resizebox{\textwidth}{!}{
\begin{tabular}{c}
\includegraphics[]{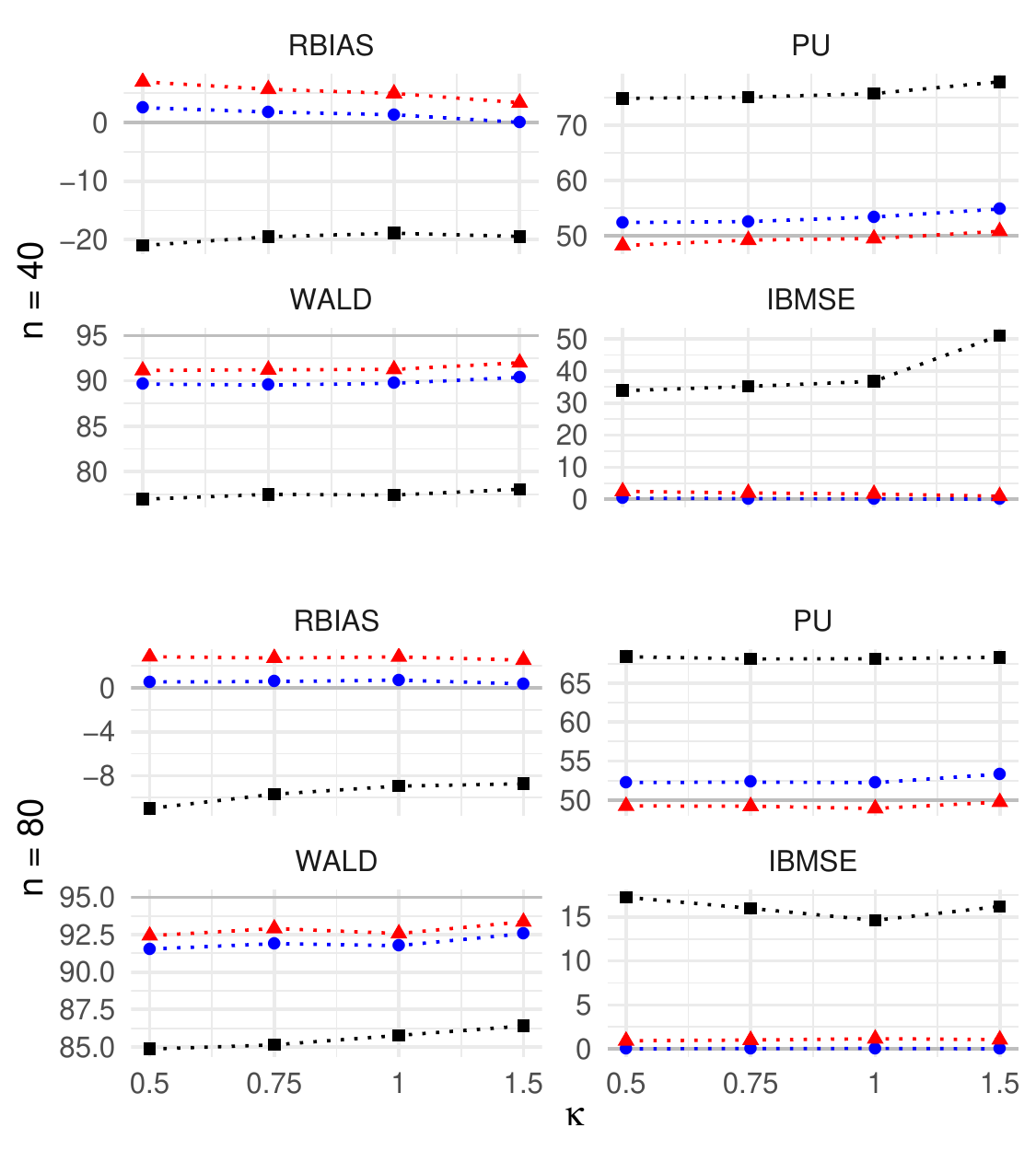}
\end{tabular}
}
\end{figure}

The summaries of the simulation results for the regression parameters when $\kappa=0.75$ are presented in Figures \ref{fig2}.
The Supplementary Material also includes 
results for $\kappa=0.5, 1, 1.5$. Figure \ref{fig3} summarizes the results for the estimators of $\kappa$.
Looking at the four performance measures, it appears that   mean and median BR  outperform the ML for small $n$. On the other hand, the results improve for all three methods as  $n$ increases.   As $\kappa$ increases,  for estimation of regression coefficients,  median BR  is comparable to mean BR  in terms of estimated relative (mean) bias, while it proves to be remarkably accurate in achieving median centering.   Moreover,  in all scenarios, median BR  provides the empirical coverages  of Wald-type confidence intervals closest to the  95\% nominal value. Finally, the results in Figure \ref{fig3} show that the improvement given by both mean and median BR over ML are substantial in all scenarios and more pronounced than in the previous case with only the intercept parameter.

\section{Case studies}
\label{sec:case}
 
We consider two case studies, namely data from an  Ames salmonella assay  and from an epileptic seizures study. 
The first data set has one explanatory variable with 6 levels and 3 observations each. 
The second data set has counts of epilectic seizures for 59 matched pairs. 

\subsection{Ames salmonella data}
Data  from an Ames salmonella reverse mutagenicity assay are presented in \citet{marg1989} and also  analysed in \citet[Table 2]{saha2005b},  \citet[Table 5]{lawl1987} and \citet{bres1984}. The response variable $Y$ corresponds to  the number of revertant colonies observed on a plate, while covariate  $x$ is the dose level of quinoline on the plate. Three observations were taken at each of six dose levels. 

%
 
 As  \citet{bres1984}, we focus on the analysis based on the 
 log-linear model
\begin{equation}\label{loglin}
\log\mu_i= \beta_0 + \beta_1 x_i+ \beta_2 \log(x_i+10),
\end{equation}
with the identity transformation for the dispersion parameter $(\phi=\kappa)$.
In the above expression, the constant 10 represents the smallest non-zero dose level. The main interest is focused on testing for mutagenic
 effect, that is $H:\beta_2 = 0$. 
\begin{table}[!ht]
\centering
\caption{Ames salmonella assay: parameter  estimates and corresponding standard errors in parenthesis.}
\label{table2}
\vspace{0.2cm}
\begin{tabular}{lcccc}
 & ML & mean BC & mean BR & median BR \\ 
  \hline
$\beta_0$ & 2.198  (0.325) & 2.210  (0.348) & 2.216 (0.352) & 2.211 (0.359) \\ 
  $\beta_1$ & -0.001  (0.00039) & -0.001 (0.00042) & -0.001  (0.00042) & -0.001  (0.00043) \\ 
  $\beta_2$ & 0.313  (0.088) & 0.311 (0.095) & 0.309 (0.096) & 0.309 (0.098) \\ 
  $\kappa $& 0.049  (0.028) & 0.063 (0.033) & 0.065 (0.033) & 0.069 (0.035) \\ 
   \hline
\end{tabular}
\end{table}

Table \ref{table2}  shows the  estimates obtained with ML, mean bias correction (BC), mean BR and
median BR. Mean and median bias reduced estimates of the dispersion parameter are pretty much comparable, but slightly different from the maximum likelihood estimate.  This is then reflected in the standard errors of the regression parameter estimates.
\begin{table}[!ht]
\centering
\caption{Ames salmonella assay: Simulation results.}
\label{table3}
\vspace{0.2cm}
\begin{tabular}{lcccc}
 & PU & RBIAS & WALD & IBMSE \\ 
  \hline
$\hat \beta_0$ & 50.95 & -0.62 & 91.77 & 0.17 \\ 
  $\tilde\beta_0$ & 49.65 & -0.12 & 93.77 & 0.01 \\ 
 $\beta_0^*$ & 49.43 & -0.05 & 93.63 & 0.00 \\ 
  $\beta_0^\dagger$ & 49.98 & -0.26 & 94.17 & 0.03 \\ \hline
  $\hat \beta_1$ & 51.49 & -1.79 & 91.56 & 0.20 \\ 
$\tilde\beta_1$  & 50.21 & -0.39 & 93.85 & 0.01 \\ 
 $\beta_1^*$ & 49.95 & -0.09 & 93.65 & 0.00 \\ 
  $\beta_1^\dagger$  & 50.14 & -0.30 & 94.14 & 0.01 \\  \hline
 $\hat\beta_2$ & 48.56 & 0.85 & 91.70 & 0.09 \\ 
   $\tilde\beta_2$ & 49.43 & 0.26 & 93.85 & 0.01 \\ 
$\beta_2^*$& 49.71 & 0.14 & 93.74 & 0.00 \\ 
   $\beta_2^\dagger$ & 49.63 & 0.21 & 94.21 & 0.01 \\ \hline
  $\hat\kappa$ & 71.88 & -22.60 & 81.07 & 20.08 \\ 
  $\tilde\kappa$ & 55.37 & 1.98 & 90.73 & 0.11 \\ 
  $\kappa^*$ & 53.71 & 3.61 & 89.08 & 0.33 \\ 
   $\kappa^\dagger$& 48.44 & 11.96 & 91.56 & 3.37 \\ 
   \hline
\end{tabular}
\end{table}

Table \ref{table3} displays the simulation results for the parameters considering
10000 replications, covariates fixed at the observed value and true
parameters set to the ML estimates based on the  observed data.  We found 641, 276 and 212 samples out of 10000 where the IWLS algorithm did not reach convergence for ML (and mean BC), mean BR  and median BR, respectively. The 641 non convergence samples are discarded for the results of ML and mean BC, while
the results of  mean and median BR   discarded  the  276 for which the IWLS did not converge for mean BR.   
         
Similarly to what was seen in the previous section, both mean and median BR are superior to ML in reducing median and mean bias of the dispersion parameter. In particular, median BR presents  empirical coverage of the 95\% Wald-type confidence intervals closest to the nominal.

\subsection{Epileptic seizures data}
We consider here the epileptic seizures data on two-week seizure counts for 59 epileptics given by  \citet{thall1990}. The data were analyzed by several authors, including \citet[Section 10.4]{venables2002} and \citet[Section 3.3]{bellio2006}. The number of seizures was recorded for a baseline period of 8 weeks, and then patients were randomly assigned to a treatment group or a control group. Counts were then recorded for four successive two-weeks periods. The response was the number of observed seizures.  We analyzed the data by comparing the response before and after the treatment, hence obtaining a set of 59 matched pairs. The only covariates in the linear predictor are then given by the two treatment indicators. We assume a negative binomial model for the response $Y_{ij}$, $i=1,\ldots,59$, $j=1,2$, with mean and variance 
\begin{equation*}
\mu_{ij}=\exp(\lambda_i+x_{ij}\beta),\quad\quad V(Y_{ij})=\mu_{ij}+\kappa \mu_{ij}^2\,, \end{equation*}
where intercepts $\lambda_i$ determine the stratified structure corresponding to each subject, $x_{i1}=(0,0)$, while $x_{i2}=(1,0)$ if subject $i$ received the placebo and $x_{i2}=(0,1)$ if  subject $i$ received the treatment.
We focus on inference about $\beta=(\beta_1,\beta_2)^\top$ and $\kappa$, while the intercepts are treated as nuisance parameters. The methods in this paper estimate anyway the whole vector of parameters. 
\begin{figure}[!ht]
\centering
\caption{Epileptic seizures: points represent the parameter estimates while the vertical lines represent 95\% Wald-type confidence intervals. }\label{fig4}
\vspace{0.15cm}
\includegraphics[scale=1]{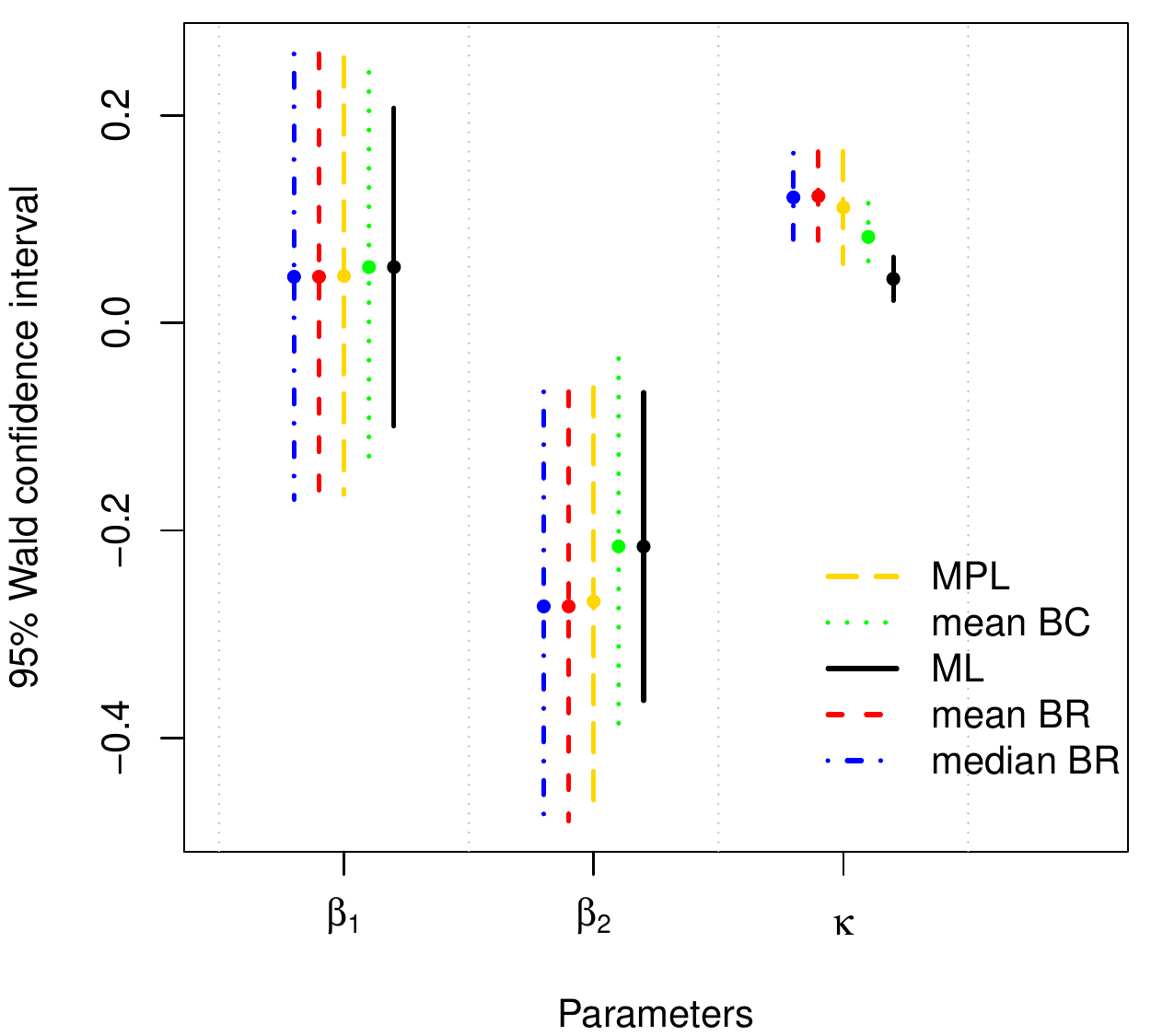}
\end{figure}
Figure \ref{fig4}  displays the parameter estimates  and the corresponding confidence intervals obtained with ML, modified profile likelihood (MPL), mean BC, mean BR and median BR. The modified profile likelihood for this model has been considered in \citet{bellio2006}.
 \begin{table}[!ht]
\centering
\caption{Epileptic seizures: Simulation results.}
\label{table4}
\vspace{0.2cm}
\begin{tabular}{lcccc}
 & PU & RBIAS & WALD & IBMSE \\ 
  \hline
 $\hat \beta_1$  & 49.80 & 0.42 & 82.22 & 0.00 \\ 
   $\tilde\beta_1$  & 50.15 & -0.16 & 89.63 & 0.00 \\ 
$\beta_1^*$ & 49.86 & 0.52 & 94.36 & 0.00 \\ 
  $\beta_1^\dagger$ & 49.89 & 0.54 & 94.40 & 0.00 \\ 
  \hline
   $\hat \beta_2$  & 49.84 & 0.09 & 82.03 & 0.00 \\ 
 $\tilde\beta_2$ & 49.45 & 0.37 & 89.04 & 0.01 \\ 
 $\beta_2^*$ & 50.63 & -0.50 & 94.55 & 0.02 \\ 
 $\beta_2^\dagger$ & 50.62 & -0.53 & 94.60 & 0.02 \\ 
 \hline
  $\hat \kappa$ & 100.00 & -79.32 & 0.39 & 3671.91 \\ 
   $\tilde\kappa$ & 93.47 & -40.99 & 40.34 & 286.52 \\ 
   $\kappa^*$& 48.78 & 3.99 & 81.44 & 1.10 \\ 
 $\kappa^\dagger$ & 48.81 & 3.89 & 82.15 & 1.08 \\ 
   \hline
\end{tabular}
\end{table}

We run 10000 replications with  covariates fixed at the observed value and true
parameters set to the observed  ML estimates.  We found  1549, 6 and 4 samples out of 10000 where the IWLS algorithm did not reach convergence for  ML (and mean BC), mean BR  and median BR, respectively. We note that, 2220 samples out of 10000 do not reach convergence using the function \texttt{glm.nb}. Hence, from  a computational point of view, the implementation in  \texttt{brnb} is more stable than that in \texttt{glm.nb}. On the other hand, it turns out that mean and median BR methods present negligible  numerical problems with respect to  ML. This is justifiable by the fact that ML tends to significantly underestimate  the dispersion parameter  producing estimates that are close to boundary of the parameter space. This is in line with the results in Table 1. The results for mean and median BR are based on samples in which the algorithm converges for both methods, while  the 1549 non convergence samples are discarded for ML and mean BC.  

Table \ref{table4} displays the results for the parameters $\beta$ and $\kappa$. For the regression coefficients, all the approaches are almost equivalent in terms of PU and RBIAS, although it should be kept in mind that the results for $\hat\beta$ and $\tilde\beta$ are based only on roughly 85\% of the samples. Concerning the dispersion parameter,   mean and median BR  outperform ML in terms of all  measures. The results  are particularly bad for ML. We also see how largely the mean squared error of $\hat \kappa$ is  affected by the huge bias in the estimator. 
We note that, in this extreme scenario, mean BC behaves similarly to ML.  In addition, both mean and median BR show  empirical coverages of  Wald-type confidence intervals remarkably close  the 95\% nominal value, while coverages for ML and mean BC are quite far from the nominal level.    

\begin{figure}[!ht]
\centering
\caption{Epileptic seizures: Simulation results for  estimators of the nuisance parameters:  ML (black solid line), mean BC (green dotted line),
 mean BR (red dashed line), and median BR (blue dot-dash line). }\label{fig5}
\vspace{0.1cm}
\resizebox{\textwidth}{!}{
\begin{tabular}{c}
\includegraphics[]{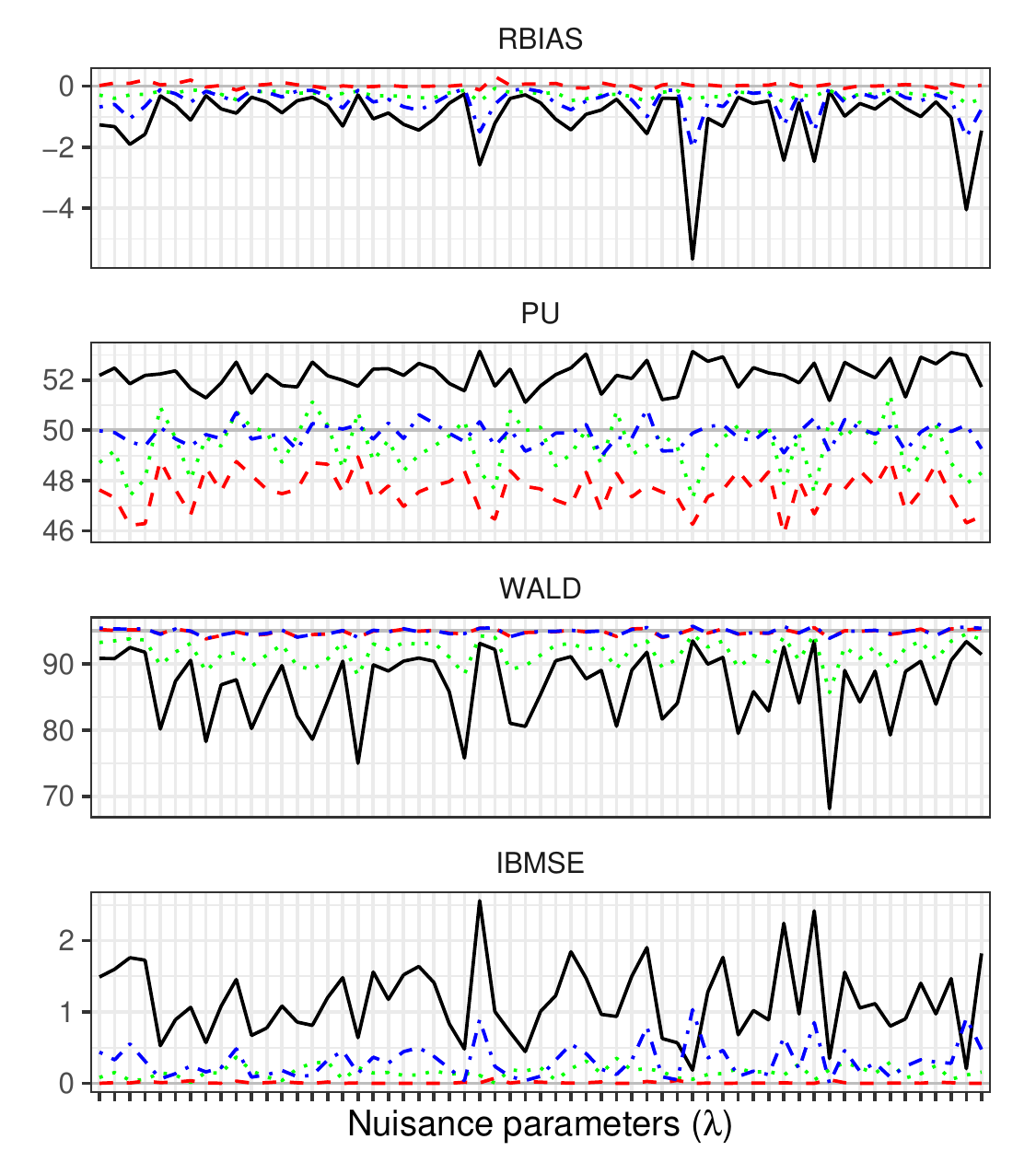}
\end{tabular}
}
\end{figure}

Although not of direct interest in the present example, both mean and median BR provide improved estimates also of the nuisance parameters. The simulation results for these are presented in Figure \ref{fig5}. Once again, we can appreciate the improved performance of 
mean and median BR by looking  at the coverages of 95\% Wald-type confidence intervals  which are closest to the nominal value.

%
%
%

\bibliographystyle{chicago}
\bibliography{nbbr.bib}

\section*{Appendix}
\label{appendix}

%


\subsection*{Quantities involved in $A^*(\theta)$ and $A^\dagger(\theta)$}

Let $\diag\{e_1,\ldots,e_n\}$ denote a diagonal matrix having $(e_1,\ldots,e_n)$ as its main diagonal. Let, in addition,  
$1_n$ be a $n$-vector of ones and $I_n$  the identity matrix of order $n$.


In order to give the expressions of matrix quantities appearing in (\ref{A*}), we will use 
the  index $s$, $s=1,\ldots,p$ for elements of $\beta$ and the subscript $\phi$ for the $\phi$ parameter.  For simplicity, the argument $\theta$ will be omitted. We get
$$
P_{\beta_s}+Q_{\beta_s}=
\left[
\begin{array}{cc}
X^\top X_s^D D^{-1}D'W X & 0_p\\
0_p^\top & 0
\end{array} 
\right] \,,
$$
where $X_s^D $ denotes the diagonal matrix with elements of the $s$th column of the matrix $X$ as its main diagonal and $D'=\diag\{d'_{1},\ldots, d'_{n}\}$. 
Moreover, letting $R=P_{\phi}+Q_{\phi}$, we have 
$$   
 R=
\left[
\begin{array}{cc}
R_{\beta\beta} & R_{\beta\phi}\\
R_{\phi\beta} & R_{\phi\phi}
\end{array}  
 \right]\,,
$$
with
\begin{align*}
R_{\beta\beta} &= \kappa'(\phi)X^\top D^2\Omega M^2 \mathcal{V}^{-2}X,\\
R_{\beta\phi} &={R}_{\phi\beta}^\top \\&= \kappa'(\phi)^2X^\top D\Omega\{M(\kappa M+I_n)\}^{-1}\{E_1-ME_2-M^3(\kappa M+I_n)^{-1}\}1_n,\\
R_{\phi\phi} &= \kappa'(\phi)^3\sum_{i=1}^n m_i\left\{ -2E(S_{3i}) + 
\frac{2\kappa^2\mu_i^3+9\kappa\mu_i^2+6\mu_i}{\kappa^3(\kappa\mu_i+1)^2}-\frac{6}{\kappa^4}\log(\kappa\mu_i+1)\right.\\
%
&\left. +2E(S_{1i}S_{2i}) -\frac{2\mu_i}{\kappa\mu_i+1}E(S_{2i}Y_i)-\frac{2\{\kappa\mu_i- (\kappa\mu_i+1)\log(\kappa\mu_i+1)\}}{\kappa^2(\kappa\mu_i+1)}E(S_{2i})\right\}  \\
&+ i_{\kappa\kappa}\kappa'(\phi)\kappa''(\phi),
\end{align*}
where 
$\Omega=\diag\{m_1,\ldots,m_n\}$,
$M=\diag\{\mu_1,\ldots,\mu_n\}$,
$\mathcal{V}=\diag\{V(Y_1),\ldots,  V(Y_n)\}$,
$S_{ai}= \sum_{j=0}^{y_i^*} j^a/(\kappa j+1)^a$,\, $a=1,2,3$, 
 $E_1=\diag\{E(S_{21}Y_1),\ldots,E(S_{2n}Y_n)\}$ and $E_2=\diag\{E(S_{21}),\ldots,E(S_{2n})\}$.\\

In order to  give the expressions for the additional quantities $u$ and $S_{\phi\phi}$ appearing in (\ref{adjglm}), we
denote by $[C]_r$  the $r$-th column of a matrix $C$ and by  $i^{ss}_{\beta\beta}$ the  $(s,s)$ element of $i_{\beta\beta}^{-1}$. Let, in addition,
$v_i=V(Y_i)$ and $v_i'=d V(Y_i)/d\mu_i=1+2\kappa \mu_i$.  Then,  $u = (u_1, \ldots, u_p)^\top$ with
\begin{align*}
	u_s = [(X^\top W X)^{-1}]_{s}^\top X^\top \left[
	\begin{array}{c}
	{h}_{s,1} \left\{d_{1} v'_1 / (6 v_1) - d'_{1}/(2 d_{1})\right\} \\
	\vdots \\
	{h}_{s,n} \left\{d_{n} v'_n / (6 v_n) - d'_{n}/(2 d_{n})\right\}
	\end{array}
	\right] \,.
\end{align*} 
In the above expression,
${h}_{s,i}$ is the
$i$th diagonal element of $X G_s X^T W$, with
$ G_s = [i_{\beta\beta}^{-1}]_{s} [i_{\beta\beta}^{-1}]_{s}^\top / ( i^{ss}_{\beta\beta})$.

Finally,
\begin{align*}
S_{\phi\phi} &= \kappa'(\phi)^3\sum_{i=1}^n m_i\left\{ -\frac{2}{3}E(S_{3i}) + \frac{1}{3}
\frac{2\kappa^2\mu_i^3+9\kappa\mu_i^2+6\mu_i}{\kappa^3(\kappa\mu_i+1)^2}-\frac{2}{\kappa^4}\log(\kappa\mu_i+1)\right.\\
&\left. +\frac{1}{2}E(S_{1i}S_{2i}) -\frac{1}{2}\frac{\mu_i}{\kappa\mu_i+1}E(S_{2i}Y_i)-\frac{\{\kappa\mu_i- (\kappa\mu_i+1)\log(\kappa\mu_i+1)\}}{2\kappa^2(\kappa\mu_i+1)}E(S_{2i})\right\}  \\
&+ \frac{1}{2}i_{\kappa\kappa}\kappa'(\phi)\kappa''(\phi)\,.
\end{align*}

\newpage
\setcounter{section}{0}
\section*{Supplementary Material for Accurate inference in negative binomial
regression}
\section{introduction}
The current report reproduces the numerical results and figures in the
main text. The outputs have been produced using R  \citep{rproject}
and the     \texttt{brnb} function available  in the forked  \texttt{brglm2}   \texttt{R}  package \citep{brglm2} on GitHub (\url{https://github.com/eulogepagui/brglm2}).

The code chunk below checks  and loads the R packages that are used for
the reproduction of numerical results in the main text.
\begin{knitrout}\footnotesize
\definecolor{shadecolor}{rgb}{0.969, 0.969, 0.969}\color{fgcolor}\begin{kframe}
\begin{alltt}
\hlkwd{library}\hlstd{(}\hlstr{"MASS"}\hlstd{)}
\hlkwd{library}\hlstd{(}\hlstr{"survival"}\hlstd{)}
\hlkwd{library}\hlstd{(}\hlstr{"ggplot2"}\hlstd{)}
\hlkwd{library}\hlstd{(}\hlstr{"gridExtra"}\hlstd{)}
\hlkwd{library}\hlstd{(}\hlstr{"cowplot"}\hlstd{)}
\end{alltt}
\end{kframe}
\end{knitrout}

We also provide code to reproduce all model fits and simulation
results in the main text. The \texttt{R} scripts to carry out the simulation
experiments, and the results from those are provided in the
\texttt{brnb\_code\_results.zip} archive. \texttt{res\_dir} is the
directory where the contents of the archive are and needs to be set
appropriately.
\begin{knitrout}\footnotesize
\definecolor{shadecolor}{rgb}{0.969, 0.969, 0.969}\color{fgcolor}\begin{kframe}
\begin{alltt}
\hlstd{res_dir} \hlkwb{<-} \hlstr{"brnb_code_results"}
\end{alltt}
\end{kframe}
\end{knitrout}

\section{Simulation studies}

 This section provides the \texttt{R} code that reproduces the numerical results
 of Section 4 of the paper.

The following code chunk uses the image file
\texttt{intercept\_simulation\_results.rda} to reproduce the reported computational diagnostics in Table 1  of the main text. 

\texttt{intercept\_simulation\_results.rda}
results by running the script\\ \texttt{brnb\_intercept\_functions.R}
which is available in the supplementary code archive.
\begin{knitrout}\footnotesize
\definecolor{shadecolor}{rgb}{0.969, 0.969, 0.969}\color{fgcolor}\begin{kframe}
\begin{alltt}
\hlkwd{load}\hlstd{(}\hlkwd{paste}\hlstd{(res_dir,} \hlstr{"intercept_simulation_results.rda"}\hlstd{,} \hlkwc{sep} \hlstd{=} \hlstr{"/"}\hlstd{))}
\hlstd{computationDiagnostic}
\end{alltt}
\begin{verbatim}
##    0.5 0.75   1 1.5  2 0.5 0.75 1 1.5 2
## A1 535  214 108  43 17  18    3 0   0 0
## A2 163   85  42  17 12   6    3 0   3 1
## A3   2    1   0   0  1   0    0 0   2 1
## A4   0    0   0   0  0   0    0 0   2 1
## A1  36    6   0   0  0   0    0 0   0 0
## A2  16    0   1   0  0   0    0 0   0 0
## A3   2    0   0   0  0   0    0 0   0 0
## A4   0    0   0   0  0   0    0 0   0 0
\end{verbatim}
\end{kframe}
\end{knitrout}

The code chunk below prepares the data for producing Figure~1 in the main text.
\begin{knitrout}\footnotesize
\definecolor{shadecolor}{rgb}{0.969, 0.969, 0.969}\color{fgcolor}\begin{kframe}
\begin{alltt}
\hlkwd{load}\hlstd{(}\hlkwd{paste}\hlstd{(res_dir,} \hlstr{"intercept_simulation_results.rda"}\hlstd{,} \hlkwc{sep} \hlstd{=} \hlstr{"/"}\hlstd{))}
\hlstd{mu} \hlkwb{<-} \hlkwd{rep}\hlstd{(}\hlkwd{c}\hlstd{(}\hlnum{2}\hlstd{,}\hlnum{5}\hlstd{),}\hlkwc{each}\hlstd{=}\hlnum{6}\hlstd{,}\hlkwc{times}\hlstd{=}\hlnum{10}\hlstd{)}
\hlstd{mu2} \hlkwb{<-} \hlkwd{as.factor}\hlstd{(mu)}
\hlkwd{levels}\hlstd{(mu2)}\hlkwb{<-} \hlkwd{c}\hlstd{(}\hlkwd{expression}\hlstd{(beta}\hlopt{*}\hlstr{" = "}\hlopt{*}\hlstd{log}\hlopt{*}\hlstr{"(2)"}\hlstd{),}
                \hlkwd{expression}\hlstd{(beta}\hlopt{*}\hlstr{" = "}\hlopt{*}\hlstd{log}\hlopt{*}\hlstr{"(5)"}\hlstd{))}
\hlstd{n} \hlkwb{<-} \hlkwd{rep}\hlstd{(}\hlkwd{c}\hlstd{(}\hlnum{20}\hlstd{,}\hlnum{50}\hlstd{),}\hlkwc{each}\hlstd{=}\hlnum{12}\hlstd{,}\hlkwc{times}\hlstd{=}\hlnum{5}\hlstd{)}
\hlstd{n2} \hlkwb{<-} \hlkwd{factor}\hlstd{(}\hlkwd{paste0}\hlstd{(}\hlstr{"n=="}\hlstd{,}\hlkwd{rep}\hlstd{(}\hlkwd{c}\hlstd{(}\hlnum{20}\hlstd{,}\hlnum{50}\hlstd{),}\hlkwc{each}\hlstd{=}\hlnum{12}\hlstd{,}\hlkwc{times}\hlstd{=}\hlnum{5}\hlstd{)),}
            \hlkwc{levels} \hlstd{=} \hlkwd{c}\hlstd{(}\hlstr{"n==20"}\hlstd{,}\hlstr{"n==50"}\hlstd{))}
\hlstd{n3} \hlkwb{<-} \hlkwd{as.factor}\hlstd{(}\hlkwd{rep}\hlstd{(}\hlkwd{c}\hlstd{(}\hlstr{"n20b"}\hlstd{,}\hlstr{"n20k"}\hlstd{,}\hlstr{"n20b"}\hlstd{,}\hlstr{"n20k"}\hlstd{,}
         \hlstr{"n50b"}\hlstd{,}\hlstr{"n50k"}\hlstd{,}\hlstr{"n50b"}\hlstd{,}\hlstr{"n50k"}\hlstd{),}\hlkwc{each}\hlstd{=}\hlnum{3}\hlstd{,}\hlkwc{times}\hlstd{=}\hlnum{5}\hlstd{))}
\hlkwd{levels}\hlstd{(n3)} \hlkwb{<-} \hlkwd{c}\hlstd{(}\hlstr{"n20b"}\hlstd{,}\hlstr{"n50b"}\hlstd{,}\hlstr{"n20k"}\hlstd{,}\hlstr{"n50k"}\hlstd{)}
\hlstd{n4} \hlkwb{<-} \hlkwd{as.factor}\hlstd{(}\hlkwd{rep}\hlstd{(}\hlkwd{c}\hlstd{(}\hlnum{1}\hlstd{,}\hlnum{3}\hlstd{,}\hlnum{1}\hlstd{,}\hlnum{3}\hlstd{,}
                   \hlnum{2}\hlstd{,}\hlnum{4}\hlstd{,}\hlnum{2}\hlstd{,}\hlnum{4}\hlstd{),}\hlkwc{each}\hlstd{=}\hlnum{3}\hlstd{,}\hlkwc{times}\hlstd{=}\hlnum{5}\hlstd{))}
\hlkwd{levels}\hlstd{(n4)} \hlkwb{<-}  \hlkwd{c}\hlstd{(}\hlstr{"n20b"}\hlstd{,}\hlstr{"n50b"}\hlstd{,}\hlstr{"n20k"}\hlstd{,}\hlstr{"n50k"}\hlstd{)}
\hlkwd{levels}\hlstd{(n4)} \hlkwb{<-} \hlkwd{c}\hlstd{(}\hlkwd{expression}\hlstd{(beta}\hlopt{*}\hlstr{", "}\hlopt{*}\hlstd{n}\hlopt{*}\hlstr{" = "}\hlopt{*}\hlstr{"20"}\hlstd{),}
               \hlkwd{expression}\hlstd{(beta}\hlopt{*}\hlstr{", "}\hlopt{*}\hlstd{n}\hlopt{*}\hlstr{" = "}\hlopt{*}\hlstr{"50"}\hlstd{),}
               \hlkwd{expression}\hlstd{(kappa}\hlopt{*}\hlstr{", "}\hlopt{*}\hlstd{n}\hlopt{*}\hlstr{" = "}\hlopt{*}\hlstr{"20"}\hlstd{),}
               \hlkwd{expression}\hlstd{(kappa}\hlopt{*}\hlstr{", "}\hlopt{*}\hlstd{n}\hlopt{*}\hlstr{" = "}\hlopt{*}\hlstr{"50"}\hlstd{))}
\hlstd{par} \hlkwb{<-} \hlkwd{rep}\hlstd{(}\hlkwd{c}\hlstd{(}\hlstr{"b"}\hlstd{,}\hlstr{"k"}\hlstd{),}\hlkwc{each}\hlstd{=}\hlnum{3}\hlstd{,}\hlkwc{times}\hlstd{=}\hlnum{20}\hlstd{)}
\hlstd{kappa} \hlkwb{<-} \hlkwd{rep}\hlstd{(}\hlkwd{c}\hlstd{(}\hlnum{1}\hlopt{:}\hlnum{5}\hlstd{),}\hlkwc{each}\hlstd{=}\hlnum{24}\hlstd{)}
\hlstd{methods} \hlkwb{<-} \hlkwd{rep}\hlstd{(}\hlkwd{c}\hlstd{(}\hlstr{"ml"}\hlstd{,}\hlstr{"br"}\hlstd{,}\hlstr{"mbr"}\hlstd{),}\hlkwc{times}\hlstd{=}\hlnum{40}\hlstd{)}
\hlstd{dataggplot} \hlkwb{<-} \hlkwd{data.frame}\hlstd{(results1DataFrame,mu,mu2,n,n2,n3,n4,par,kappa,methods)}
\end{alltt}
\end{kframe}
\end{knitrout}
Figure~1 is the result of
\begin{knitrout}\footnotesize
\definecolor{shadecolor}{rgb}{0.969, 0.969, 0.969}\color{fgcolor}\begin{kframe}
\begin{alltt}
\hlcom{## Relative bias; probability of underestimation and coverages }
\hlstd{p1}  \hlkwb{<-} \hlkwd{ggplot}\hlstd{(dataggplot,}\hlkwd{aes}\hlstd{(}\hlkwc{x} \hlstd{= kappa,} \hlkwc{y} \hlstd{= rbias))} \hlopt{+}
  \hlkwd{geom_hline}\hlstd{(}\hlkwd{aes}\hlstd{(}\hlkwc{yintercept} \hlstd{=} \hlnum{0}\hlstd{),} \hlkwc{col} \hlstd{=} \hlstr{"grey"}\hlstd{)} \hlopt{+}
  \hlkwd{geom_line}\hlstd{(}\hlkwd{aes}\hlstd{(}\hlkwc{linetype}\hlstd{=methods,}\hlkwc{colour}\hlstd{=methods ))}\hlopt{+}
  \hlkwd{geom_point}\hlstd{(}\hlkwd{aes}\hlstd{(}\hlkwc{shape}\hlstd{=methods,}\hlkwc{colour}\hlstd{=methods)}
  \hlstd{)} \hlopt{+}
  \hlkwd{labs}\hlstd{(}\hlkwc{x} \hlstd{=} \hlstr{""}\hlstd{,} \hlkwc{y} \hlstd{=} \hlstr{"RBIAS"}\hlstd{)} \hlopt{+}
  \hlkwd{facet_grid}\hlstd{(mu2} \hlopt{~} \hlstd{n4,} \hlkwc{labeller} \hlstd{= label_parsed}
  \hlstd{)}\hlopt{+}
  \hlkwd{scale_colour_manual}\hlstd{(}\hlkwc{values}\hlstd{=}\hlkwd{c}\hlstd{(}\hlstr{"blue"}\hlstd{,}\hlstr{"red"}\hlstd{,} \hlstr{"black"}\hlstd{))}\hlopt{+}
  \hlkwd{scale_linetype_manual}\hlstd{(}\hlkwc{values} \hlstd{=} \hlkwd{c}\hlstd{(}\hlstr{"dotted"}\hlstd{,}\hlstr{"dotted"}\hlstd{,}\hlstr{"dotted"}\hlstd{))}\hlopt{+}
  \hlkwd{scale_x_discrete}\hlstd{(}\hlstr{""}\hlstd{,}\hlkwc{limits}\hlstd{=}\hlkwd{factor}\hlstd{(}\hlkwd{c}\hlstd{(}\hlnum{1}\hlopt{:}\hlnum{5}\hlstd{)),} \hlkwc{breaks}\hlstd{=}\hlkwd{c}\hlstd{(}\hlnum{1}\hlopt{:}\hlnum{5}\hlstd{),}
                   \hlkwc{labels}\hlstd{=}\hlkwd{c}\hlstd{(}\hlnum{0.5}\hlstd{,}\hlnum{0.75}\hlstd{,}\hlnum{1}\hlstd{,}\hlnum{1.5}\hlstd{,}\hlnum{2}\hlstd{))}\hlopt{+}
  \hlkwd{theme_bw}\hlstd{()}\hlopt{+}
  \hlkwd{theme}\hlstd{(}\hlkwc{legend.position} \hlstd{=} \hlstr{""}\hlstd{,}\hlkwc{strip.background} \hlstd{=} \hlkwd{element_blank}\hlstd{(),}
        \hlkwc{axis.ticks.x}\hlstd{=}\hlkwd{element_blank}\hlstd{(),}\hlkwc{axis.text.x} \hlstd{=} \hlkwd{element_blank}\hlstd{(),}
        \hlkwc{plot.margin}\hlstd{=}\hlkwd{unit}\hlstd{(}\hlkwd{c}\hlstd{(}\hlnum{0}\hlstd{,}\hlnum{0}\hlstd{,}\hlopt{-}\hlnum{0.1}\hlstd{,}\hlnum{0.03}\hlstd{),} \hlstr{"cm"}\hlstd{)}
  \hlstd{)}
\hlstd{p2} \hlkwb{<-} \hlkwd{ggplot}\hlstd{(dataggplot,}\hlkwd{aes}\hlstd{(}\hlkwc{x} \hlstd{= kappa,} \hlkwc{y} \hlstd{= pu))} \hlopt{+}
  \hlkwd{geom_hline}\hlstd{(}\hlkwd{aes}\hlstd{(}\hlkwc{yintercept} \hlstd{=} \hlnum{50}\hlstd{),} \hlkwc{col} \hlstd{=} \hlstr{"grey"}\hlstd{)} \hlopt{+}
  \hlkwd{geom_line}\hlstd{(}\hlkwd{aes}\hlstd{(}\hlkwc{linetype}\hlstd{=methods,}\hlkwc{colour}\hlstd{=methods ))}\hlopt{+}
  \hlkwd{geom_point}\hlstd{(}\hlkwd{aes}\hlstd{(}\hlkwc{shape}\hlstd{=methods,}\hlkwc{colour}\hlstd{=methods)}
  \hlstd{)} \hlopt{+}
  \hlkwd{labs}\hlstd{(}\hlkwc{x} \hlstd{=} \hlstr{""}\hlstd{,} \hlkwc{y} \hlstd{=} \hlstr{"PU"}\hlstd{)} \hlopt{+}
  \hlkwd{facet_grid}\hlstd{(mu2} \hlopt{~} \hlstd{n4,} \hlkwc{labeller} \hlstd{= label_parsed}
  \hlstd{)}\hlopt{+}
  \hlkwd{scale_colour_manual}\hlstd{(}\hlkwc{values}\hlstd{=}\hlkwd{c}\hlstd{(}\hlstr{"blue"}\hlstd{,}\hlstr{"red"}\hlstd{,} \hlstr{"black"}\hlstd{))}\hlopt{+}
  \hlkwd{scale_linetype_manual}\hlstd{(}\hlkwc{values} \hlstd{=} \hlkwd{c}\hlstd{(}\hlstr{"dotted"}\hlstd{,}\hlstr{"dotted"}\hlstd{,}\hlstr{"dotted"}\hlstd{))}\hlopt{+}
  \hlkwd{scale_x_discrete}\hlstd{(}\hlstr{""}\hlstd{,}\hlkwc{limits}\hlstd{=}\hlkwd{factor}\hlstd{(}\hlkwd{c}\hlstd{(}\hlnum{1}\hlopt{:}\hlnum{5}\hlstd{)),} \hlkwc{breaks}\hlstd{=}\hlkwd{c}\hlstd{(}\hlnum{1}\hlopt{:}\hlnum{5}\hlstd{),}
                   \hlkwc{labels}\hlstd{=}\hlkwd{c}\hlstd{(}\hlnum{0.5}\hlstd{,}\hlnum{0.75}\hlstd{,}\hlnum{1}\hlstd{,}\hlnum{1.5}\hlstd{,}\hlnum{2}\hlstd{))}\hlopt{+}
  \hlkwd{theme_bw}\hlstd{()}\hlopt{+}
  \hlkwd{theme}\hlstd{(}\hlkwc{legend.position} \hlstd{=} \hlstr{""}\hlstd{,}\hlkwc{strip.background} \hlstd{=} \hlkwd{element_blank}\hlstd{(),}
        \hlkwc{strip.text.x} \hlstd{=} \hlkwd{element_blank}\hlstd{(),}
        \hlkwc{axis.text.x} \hlstd{=} \hlkwd{element_blank}\hlstd{(),}
        \hlkwc{axis.ticks.x}\hlstd{=}\hlkwd{element_blank}\hlstd{(),}
        \hlkwc{plot.margin}\hlstd{=}\hlkwd{unit}\hlstd{(}\hlkwd{c}\hlstd{(}\hlopt{-}\hlnum{0.1}\hlstd{,}\hlnum{0}\hlstd{,}\hlopt{-}\hlnum{0.1}\hlstd{,}\hlnum{0.03}\hlstd{),} \hlstr{"cm"}\hlstd{))}
\hlstd{p3} \hlkwb{<-} \hlkwd{ggplot}\hlstd{(dataggplot,}\hlkwd{aes}\hlstd{(}\hlkwc{x} \hlstd{= kappa,} \hlkwc{y} \hlstd{= cov))} \hlopt{+}
  \hlkwd{geom_hline}\hlstd{(}\hlkwd{aes}\hlstd{(}\hlkwc{yintercept} \hlstd{=} \hlnum{95}\hlstd{),} \hlkwc{col} \hlstd{=} \hlstr{"grey"}\hlstd{)} \hlopt{+}
  \hlkwd{geom_line}\hlstd{(}\hlkwd{aes}\hlstd{(}\hlkwc{linetype}\hlstd{=methods,}\hlkwc{colour}\hlstd{=methods ))}\hlopt{+}
  \hlkwd{geom_point}\hlstd{(}\hlkwd{aes}\hlstd{(}\hlkwc{shape}\hlstd{=methods,}\hlkwc{colour}\hlstd{=methods)}
  \hlstd{)} \hlopt{+}
  \hlkwd{labs}\hlstd{(}\hlkwc{x} \hlstd{=} \hlstr{""}\hlstd{,} \hlkwc{y} \hlstd{=} \hlstr{" WALD"}\hlstd{)} \hlopt{+}
  \hlkwd{facet_grid}\hlstd{(mu2} \hlopt{~} \hlstd{n4,} \hlkwc{labeller} \hlstd{= label_parsed}
  \hlstd{)}\hlopt{+}
  \hlkwd{scale_colour_manual}\hlstd{(}\hlkwc{values}\hlstd{=}\hlkwd{c}\hlstd{(}\hlstr{"blue"}\hlstd{,}\hlstr{"red"}\hlstd{,} \hlstr{"black"}\hlstd{))}\hlopt{+}
  \hlkwd{scale_linetype_manual}\hlstd{(}\hlkwc{values} \hlstd{=} \hlkwd{c}\hlstd{(}\hlstr{"dotted"}\hlstd{,}\hlstr{"dotted"}\hlstd{,}\hlstr{"dotted"}\hlstd{))}\hlopt{+}
  \hlkwd{scale_x_discrete}\hlstd{(}\hlkwd{expression}\hlstd{(kappa),}\hlkwc{limits}\hlstd{=}\hlkwd{factor}\hlstd{(}\hlkwd{c}\hlstd{(}\hlnum{1}\hlopt{:}\hlnum{5}\hlstd{)),} \hlkwc{breaks}\hlstd{=}\hlkwd{c}\hlstd{(}\hlnum{1}\hlopt{:}\hlnum{5}\hlstd{),}
                   \hlkwc{labels}\hlstd{=}\hlkwd{c}\hlstd{(}\hlnum{0.5}\hlstd{,}\hlnum{0.75}\hlstd{,}\hlnum{1}\hlstd{,}\hlnum{1.5}\hlstd{,}\hlnum{2}\hlstd{))}\hlopt{+}
  \hlkwd{theme_bw}\hlstd{()}\hlopt{+}
  \hlkwd{theme}\hlstd{(}\hlkwc{legend.position} \hlstd{=} \hlstr{""}\hlstd{,}\hlkwc{strip.background} \hlstd{=} \hlkwd{element_blank}\hlstd{(),}
        \hlkwc{strip.text.x} \hlstd{=} \hlkwd{element_blank}\hlstd{(),}
        \hlkwc{plot.margin}\hlstd{=}\hlkwd{unit}\hlstd{(}\hlkwd{c}\hlstd{(}\hlopt{-}\hlnum{0.1}\hlstd{,}\hlnum{0}\hlstd{,}\hlnum{0}\hlstd{,}\hlnum{0.03}\hlstd{),} \hlstr{"cm"}\hlstd{))}
\end{alltt}
\end{kframe}
\end{knitrout}
\begin{knitrout}\footnotesize
\definecolor{shadecolor}{rgb}{0.969, 0.969, 0.969}\color{fgcolor}\begin{kframe}
\begin{alltt}
  \hlkwd{plot_grid}\hlstd{(p1, p2, p3,} \hlkwc{labels}\hlstd{=}\hlkwd{c}\hlstd{(}\hlstr{""}\hlstd{,} \hlstr{""}\hlstd{,} \hlstr{""}\hlstd{),} \hlkwc{ncol} \hlstd{=} \hlnum{1}\hlstd{,} \hlkwc{nrow} \hlstd{=} \hlnum{3}\hlstd{,}
          \hlkwc{align} \hlstd{=} \hlstr{"v"}\hlstd{)}
\end{alltt}
\end{kframe}\begin{figure}
\includegraphics[width=\maxwidth]{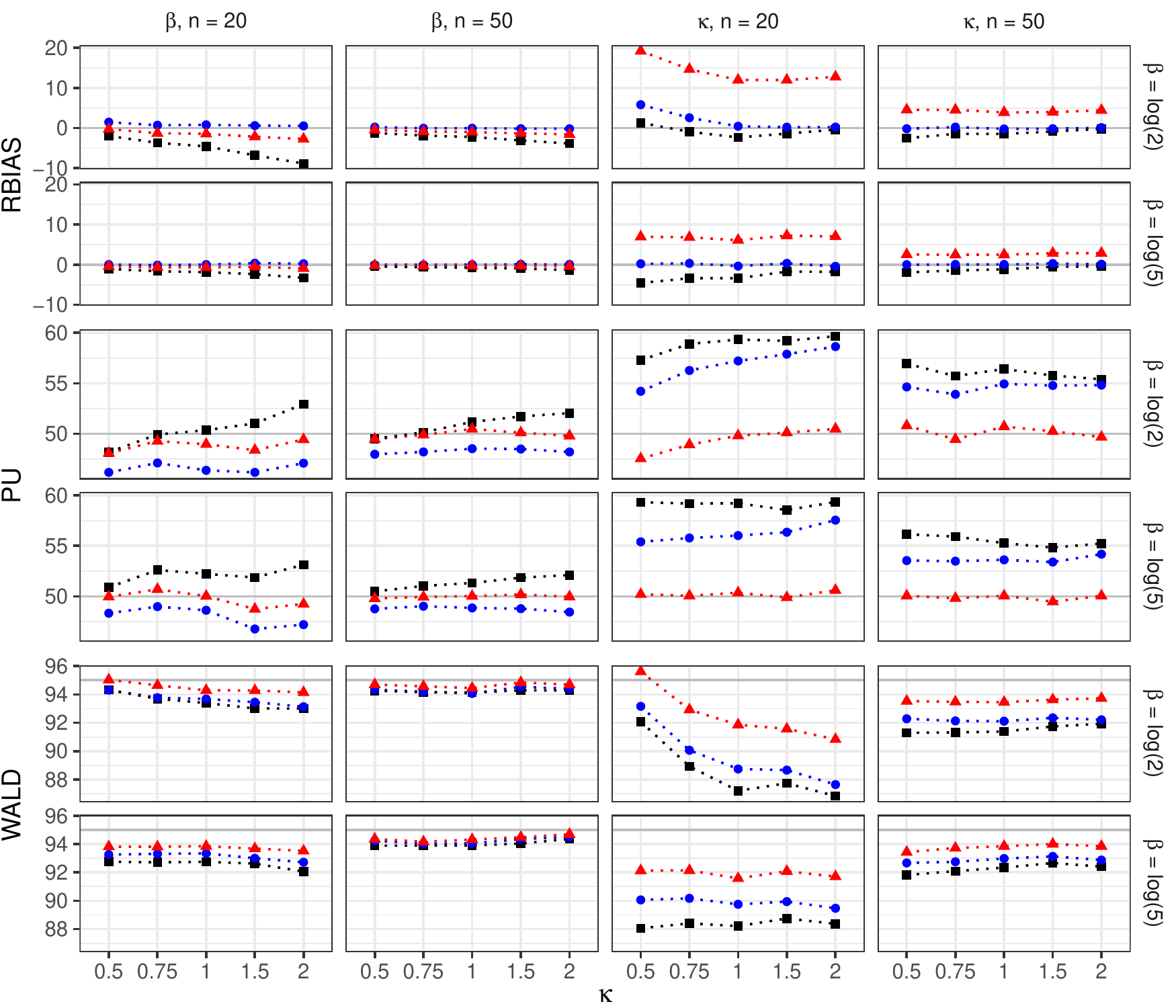} \caption[Estimated relative bias (RBIAS), probability of under estimation (PU) and  coverage probability of 95\% Wald-type confidence intervals (WALD) for the intercept $\beta=\log\mu$]{Estimated relative bias (RBIAS), probability of under estimation (PU) and  coverage probability of 95\% Wald-type confidence intervals (WALD) for the intercept $\beta=\log\mu$. Results for ML (black squares), mean BR (blue circles) and median BR (red triangles}\label{fig:fig1b}
\end{figure}

\end{knitrout}

The code chunk below prepares the data and ggplot objects for producing Figure~2, Figure~3, Figure~4, Figure~5, Figure~6  in the main text. 
\begin{knitrout}\footnotesize
\definecolor{shadecolor}{rgb}{0.969, 0.969, 0.969}\color{fgcolor}\begin{kframe}
\begin{alltt}
\hlkwd{load}\hlstd{(}\hlkwd{paste}\hlstd{(res_dir,} \hlstr{"covariates_simulation_results.rda"}\hlstd{,} \hlkwc{sep} \hlstd{=} \hlstr{"/"}\hlstd{))}
\hlstd{coef} \hlkwb{<-} \hlkwd{rep}\hlstd{(}\hlkwd{rep}\hlstd{(}\hlnum{1}\hlopt{:}\hlnum{6}\hlstd{,}\hlkwc{each}\hlstd{=}\hlnum{3}\hlstd{),}\hlnum{20}\hlstd{)}
\hlstd{measures} \hlkwb{<-} \hlkwd{as.factor}\hlstd{(}\hlkwd{rep}\hlstd{(}\hlkwd{rep}\hlstd{(}\hlkwd{c}\hlstd{(}\hlnum{1}\hlstd{,}\hlnum{2}\hlstd{,}\hlnum{3}\hlstd{,}\hlnum{4}\hlstd{),}\hlkwc{each}\hlstd{=}\hlnum{18}\hlstd{),}\hlnum{5}\hlstd{))}
\hlkwd{levels}\hlstd{(measures)} \hlkwb{<-} \hlkwd{c}\hlstd{(}\hlstr{"RBIAS"}\hlstd{,}\hlstr{"PU"}\hlstd{,}\hlstr{"WALD"}\hlstd{,}\hlstr{"IBMSE"}\hlstd{)}
\hlstd{methods} \hlkwb{<-} \hlkwd{rep}\hlstd{(}\hlkwd{c}\hlstd{(}\hlstr{"ml"}\hlstd{,}\hlstr{"br"}\hlstd{,}\hlstr{"mbr"}\hlstd{),}\hlnum{120}\hlstd{)}
\hlstd{avalues} \hlkwb{<-} \hlkwd{rep}\hlstd{(}\hlnum{1}\hlopt{:}\hlnum{5}\hlstd{,}\hlkwc{each}\hlstd{=}\hlnum{72}\hlstd{)}
\hlstd{dataggplot} \hlkwb{<-} \hlkwd{data.frame}\hlstd{(yvalues,methods,avalues,measures,coef)}
\hlstd{dataggplot2} \hlkwb{<-} \hlkwd{data.frame}\hlstd{(yvalues2,methods,avalues,measures,coef)}
\hlstd{hline_dat} \hlkwb{<-} \hlkwd{data.frame}\hlstd{(}\hlkwc{measures}\hlstd{=}\hlkwd{c}\hlstd{(}\hlstr{"RBIAS"} \hlstd{,}  \hlstr{"PU"}\hlstd{,}   \hlstr{"WALD"}\hlstd{,} \hlstr{"IBMSE"}\hlstd{),}
                     \hlkwc{threshold}\hlstd{=}\hlkwd{c}\hlstd{(}\hlnum{0}\hlstd{,} \hlnum{50}\hlstd{,} \hlnum{95}\hlstd{,} \hlnum{0}\hlstd{))}
\hlstd{coef} \hlkwb{<-} \hlkwd{rep}\hlstd{(}\hlkwd{rep}\hlstd{(}\hlnum{1}\hlopt{:}\hlnum{6}\hlstd{,}\hlkwc{each}\hlstd{=}\hlnum{3}\hlstd{),}\hlnum{20}\hlstd{)}
\hlstd{measures} \hlkwb{<-} \hlkwd{as.factor}\hlstd{(}\hlkwd{rep}\hlstd{(}\hlkwd{rep}\hlstd{(}\hlkwd{c}\hlstd{(}\hlnum{1}\hlstd{,}\hlnum{2}\hlstd{,}\hlnum{3}\hlstd{,}\hlnum{4}\hlstd{),}\hlkwc{each}\hlstd{=}\hlnum{18}\hlstd{),}\hlnum{5}\hlstd{))}
\hlkwd{levels}\hlstd{(measures)} \hlkwb{<-} \hlkwd{c}\hlstd{(}\hlstr{"RBIAS"}\hlstd{,}\hlstr{"PU"}\hlstd{,}\hlstr{"WALD"}\hlstd{,}\hlstr{"IBMSE"}\hlstd{)}
\hlstd{methods} \hlkwb{<-} \hlkwd{rep}\hlstd{(}\hlkwd{c}\hlstd{(}\hlstr{"ml"}\hlstd{,}\hlstr{"br"}\hlstd{,}\hlstr{"mbr"}\hlstd{),}\hlnum{120}\hlstd{)}
\hlstd{avalues} \hlkwb{<-} \hlkwd{rep}\hlstd{(}\hlnum{1}\hlopt{:}\hlnum{5}\hlstd{,}\hlkwc{each}\hlstd{=}\hlnum{72}\hlstd{)}
\hlstd{dataggplot} \hlkwb{<-} \hlkwd{data.frame}\hlstd{(yvalues,methods,avalues,measures,coef)}
\hlstd{dataggplot2} \hlkwb{<-} \hlkwd{data.frame}\hlstd{(yvalues2,methods,avalues,measures,coef)}
\hlstd{hline_dat} \hlkwb{<-} \hlkwd{data.frame}\hlstd{(}\hlkwc{measures}\hlstd{=}\hlkwd{c}\hlstd{(}\hlstr{"RBIAS"} \hlstd{,}  \hlstr{"PU"}\hlstd{,}   \hlstr{"WALD"}\hlstd{,} \hlstr{"IBMSE"}\hlstd{),}
                     \hlkwc{threshold}\hlstd{=}\hlkwd{c}\hlstd{(}\hlnum{0}\hlstd{,} \hlnum{50}\hlstd{,} \hlnum{95}\hlstd{,} \hlnum{0}\hlstd{))}
\hlcom{## n = 40 ##}
\hlstd{p0.5} \hlkwb{<-} \hlkwd{ggplot}\hlstd{(}\hlkwd{subset}\hlstd{(dataggplot,(avalues}\hlopt{==}\hlnum{2} \hlopt{&} \hlstd{coef}\hlopt{!=}\hlnum{6}\hlstd{)),}\hlkwd{aes}\hlstd{(}\hlkwc{x} \hlstd{= coef,} \hlkwc{y} \hlstd{= yvalues))} \hlopt{+}
  \hlkwd{geom_hline}\hlstd{(}\hlkwc{data}\hlstd{=hline_dat,} \hlkwd{aes}\hlstd{(}\hlkwc{yintercept}\hlstd{=threshold),} \hlkwc{col} \hlstd{=} \hlstr{"grey"}\hlstd{)} \hlopt{+}
  \hlkwd{geom_line}\hlstd{(}\hlkwd{aes}\hlstd{(}\hlkwc{linetype}\hlstd{=methods,}\hlkwc{colour}\hlstd{=methods ))}\hlopt{+}
  \hlkwd{geom_point}\hlstd{(}\hlkwd{aes}\hlstd{(}\hlkwc{shape}\hlstd{=methods,}\hlkwc{colour}\hlstd{=methods))} \hlopt{+}
  \hlkwd{labs}\hlstd{(}\hlkwc{x} \hlstd{=} \hlstr{""}\hlstd{,} \hlkwc{y} \hlstd{=} \hlstr{"n = 40"}\hlstd{)} \hlopt{+}
  \hlkwd{facet_wrap}\hlstd{(} \hlopt{~} \hlstd{measures ,}\hlkwc{scales} \hlstd{=} \hlstr{"free_y"}\hlstd{,}\hlkwc{labeller} \hlstd{= label_parsed)}\hlopt{+}
  \hlkwd{scale_colour_manual}\hlstd{(}\hlkwc{values}\hlstd{=}\hlkwd{c}\hlstd{(}\hlstr{"blue"}\hlstd{,}\hlstr{"red"}\hlstd{,} \hlstr{"black"}\hlstd{))}\hlopt{+}
  \hlkwd{scale_linetype_manual}\hlstd{(}\hlkwc{values} \hlstd{=} \hlkwd{c}\hlstd{(}\hlstr{"dotted"}\hlstd{,}\hlstr{"dotted"}\hlstd{,}\hlstr{"dotted"}\hlstd{))}\hlopt{+}
  \hlkwd{scale_x_continuous}\hlstd{(}\hlkwc{labels}\hlstd{=}\hlkwd{c}\hlstd{(}\hlkwd{expression}\hlstd{(beta[}\hlnum{1}\hlstd{]),}\hlkwd{expression}\hlstd{(beta[}\hlnum{2}\hlstd{]),}
             \hlkwd{expression}\hlstd{(beta[}\hlnum{3}\hlstd{]),}\hlkwd{expression}\hlstd{(beta[}\hlnum{4}\hlstd{]),}
             \hlkwd{expression}\hlstd{(beta[}\hlnum{5}\hlstd{])))}\hlopt{+}
  \hlkwd{theme_minimal}\hlstd{()} \hlopt{+}
  \hlkwd{theme}\hlstd{(}\hlkwc{legend.position} \hlstd{=} \hlstr{"none"}\hlstd{,}
        \hlkwc{plot.title} \hlstd{=} \hlkwd{element_text}\hlstd{(}\hlkwc{hjust} \hlstd{=} \hlnum{0.5}\hlstd{),}
        \hlkwc{axis.text.x} \hlstd{=} \hlkwd{element_blank}\hlstd{(),}
        \hlkwc{axis.ticks.x}\hlstd{=}\hlkwd{element_blank}\hlstd{(),}
        \hlkwc{plot.margin}\hlstd{=}\hlkwd{unit}\hlstd{(}\hlkwd{c}\hlstd{(}\hlnum{0}\hlstd{,}\hlnum{0}\hlstd{,}\hlopt{-}\hlnum{0.1}\hlstd{,}\hlnum{0.03}\hlstd{),} \hlstr{"cm"}\hlstd{))}

\hlstd{p0.75} \hlkwb{<-} \hlkwd{ggplot}\hlstd{(}\hlkwd{subset}\hlstd{(dataggplot,(avalues}\hlopt{==}\hlnum{3} \hlopt{&} \hlstd{coef}\hlopt{!=}\hlnum{6}\hlstd{)),}\hlkwd{aes}\hlstd{(}\hlkwc{x} \hlstd{= coef,} \hlkwc{y} \hlstd{= yvalues))} \hlopt{+}
  \hlkwd{geom_hline}\hlstd{(}\hlkwc{data}\hlstd{=hline_dat,} \hlkwd{aes}\hlstd{(}\hlkwc{yintercept}\hlstd{=threshold),} \hlkwc{col} \hlstd{=} \hlstr{"grey"}\hlstd{)} \hlopt{+}
  \hlkwd{geom_line}\hlstd{(}\hlkwd{aes}\hlstd{(}\hlkwc{linetype}\hlstd{=methods,}\hlkwc{colour}\hlstd{=methods ))}\hlopt{+}
  \hlkwd{geom_point}\hlstd{(}\hlkwd{aes}\hlstd{(}\hlkwc{shape}\hlstd{=methods,}\hlkwc{colour}\hlstd{=methods))} \hlopt{+}
  \hlkwd{labs}\hlstd{(}\hlkwc{x} \hlstd{=} \hlstr{""}\hlstd{,} \hlkwc{y} \hlstd{=} \hlstr{"n = 40"}\hlstd{)} \hlopt{+}
  \hlkwd{facet_wrap}\hlstd{(} \hlopt{~} \hlstd{measures ,}\hlkwc{scales} \hlstd{=} \hlstr{"free_y"}\hlstd{,}\hlkwc{labeller} \hlstd{= label_parsed)}\hlopt{+}
  \hlkwd{scale_colour_manual}\hlstd{(}\hlkwc{values}\hlstd{=}\hlkwd{c}\hlstd{(}\hlstr{"blue"}\hlstd{,}\hlstr{"red"}\hlstd{,} \hlstr{"black"}\hlstd{))}\hlopt{+}
  \hlkwd{scale_linetype_manual}\hlstd{(}\hlkwc{values} \hlstd{=} \hlkwd{c}\hlstd{(}\hlstr{"dotted"}\hlstd{,}\hlstr{"dotted"}\hlstd{,}\hlstr{"dotted"}\hlstd{))}\hlopt{+}
  \hlkwd{scale_x_continuous}\hlstd{(}\hlstr{""}\hlstd{,}\hlkwc{labels}\hlstd{=}\hlkwd{c}\hlstd{(}\hlkwd{expression}\hlstd{(beta[}\hlnum{1}\hlstd{]),}\hlkwd{expression}\hlstd{(beta[}\hlnum{2}\hlstd{]),}
                            \hlkwd{expression}\hlstd{(beta[}\hlnum{3}\hlstd{]),}\hlkwd{expression}\hlstd{(beta[}\hlnum{4}\hlstd{]),}
                            \hlkwd{expression}\hlstd{(beta[}\hlnum{5}\hlstd{])))}\hlopt{+}
  \hlkwd{theme_minimal}\hlstd{()} \hlopt{+}
  \hlkwd{theme}\hlstd{(}\hlkwc{legend.position} \hlstd{=} \hlstr{"none"}\hlstd{,}
        \hlkwc{plot.title} \hlstd{=} \hlkwd{element_text}\hlstd{(}\hlkwc{hjust} \hlstd{=} \hlnum{0.5}\hlstd{),}
        \hlkwc{axis.text.x} \hlstd{=} \hlkwd{element_blank}\hlstd{(),}
        \hlkwc{axis.ticks.x}\hlstd{=}\hlkwd{element_blank}\hlstd{(),}
        \hlkwc{plot.margin}\hlstd{=}\hlkwd{unit}\hlstd{(}\hlkwd{c}\hlstd{(}\hlnum{0}\hlstd{,}\hlnum{0}\hlstd{,}\hlopt{-}\hlnum{0.1}\hlstd{,}\hlnum{0.03}\hlstd{),} \hlstr{"cm"}\hlstd{))}

\hlstd{p1} \hlkwb{<-} \hlkwd{ggplot}\hlstd{(}\hlkwd{subset}\hlstd{(dataggplot,(avalues}\hlopt{==}\hlnum{4} \hlopt{&} \hlstd{coef}\hlopt{!=}\hlnum{6}\hlstd{)),}\hlkwd{aes}\hlstd{(}\hlkwc{x} \hlstd{= coef,} \hlkwc{y} \hlstd{= yvalues))} \hlopt{+}
  \hlkwd{geom_hline}\hlstd{(}\hlkwc{data}\hlstd{=hline_dat,} \hlkwd{aes}\hlstd{(}\hlkwc{yintercept}\hlstd{=threshold),} \hlkwc{col} \hlstd{=} \hlstr{"grey"}\hlstd{)} \hlopt{+}
  \hlkwd{geom_line}\hlstd{(}\hlkwd{aes}\hlstd{(}\hlkwc{linetype}\hlstd{=methods,}\hlkwc{colour}\hlstd{=methods ))}\hlopt{+}
  \hlkwd{geom_point}\hlstd{(}\hlkwd{aes}\hlstd{(}\hlkwc{shape}\hlstd{=methods,}\hlkwc{colour}\hlstd{=methods))} \hlopt{+}
  \hlkwd{labs}\hlstd{(}\hlkwc{x} \hlstd{=} \hlstr{""}\hlstd{,} \hlkwc{y} \hlstd{=} \hlstr{"n = 40"}\hlstd{)} \hlopt{+}
  \hlkwd{facet_wrap}\hlstd{(} \hlopt{~} \hlstd{measures ,}\hlkwc{scales} \hlstd{=} \hlstr{"free_y"}\hlstd{,}\hlkwc{labeller} \hlstd{= label_parsed)}\hlopt{+}
  \hlkwd{scale_colour_manual}\hlstd{(}\hlkwc{values}\hlstd{=}\hlkwd{c}\hlstd{(}\hlstr{"blue"}\hlstd{,}\hlstr{"red"}\hlstd{,} \hlstr{"black"}\hlstd{))}\hlopt{+}
  \hlkwd{scale_linetype_manual}\hlstd{(}\hlkwc{values} \hlstd{=} \hlkwd{c}\hlstd{(}\hlstr{"dotted"}\hlstd{,}\hlstr{"dotted"}\hlstd{,}\hlstr{"dotted"}\hlstd{))}\hlopt{+}
  \hlkwd{scale_x_continuous}\hlstd{(}\hlstr{""}\hlstd{,} \hlkwc{labels}\hlstd{=}\hlkwd{c}\hlstd{(}\hlkwd{expression}\hlstd{(beta[}\hlnum{1}\hlstd{]),}\hlkwd{expression}\hlstd{(beta[}\hlnum{2}\hlstd{]),}
                            \hlkwd{expression}\hlstd{(beta[}\hlnum{3}\hlstd{]),}\hlkwd{expression}\hlstd{(beta[}\hlnum{4}\hlstd{]),}
                            \hlkwd{expression}\hlstd{(beta[}\hlnum{5}\hlstd{])))}\hlopt{+}
  \hlkwd{theme_minimal}\hlstd{()} \hlopt{+}
  \hlkwd{theme}\hlstd{(}\hlkwc{legend.position} \hlstd{=} \hlstr{"none"}\hlstd{,}
        \hlkwc{plot.title} \hlstd{=} \hlkwd{element_text}\hlstd{(}\hlkwc{hjust} \hlstd{=} \hlnum{0.5}\hlstd{),}
        \hlkwc{axis.text.x} \hlstd{=} \hlkwd{element_blank}\hlstd{(),}
        \hlkwc{axis.ticks.x}\hlstd{=}\hlkwd{element_blank}\hlstd{(),}
        \hlkwc{plot.margin}\hlstd{=}\hlkwd{unit}\hlstd{(}\hlkwd{c}\hlstd{(}\hlnum{0}\hlstd{,}\hlnum{0}\hlstd{,}\hlopt{-}\hlnum{0.1}\hlstd{,}\hlnum{0.03}\hlstd{),} \hlstr{"cm"}\hlstd{))}

\hlstd{p1.5} \hlkwb{<-} \hlkwd{ggplot}\hlstd{(}\hlkwd{subset}\hlstd{(dataggplot,(avalues}\hlopt{==}\hlnum{5} \hlopt{&} \hlstd{coef}\hlopt{!=}\hlnum{6}\hlstd{)),}\hlkwd{aes}\hlstd{(}\hlkwc{x} \hlstd{= coef,} \hlkwc{y} \hlstd{= yvalues))} \hlopt{+}
  \hlkwd{geom_hline}\hlstd{(}\hlkwc{data}\hlstd{=hline_dat,} \hlkwd{aes}\hlstd{(}\hlkwc{yintercept}\hlstd{=threshold),} \hlkwc{col} \hlstd{=} \hlstr{"grey"}\hlstd{)} \hlopt{+}
  \hlkwd{geom_line}\hlstd{(}\hlkwd{aes}\hlstd{(}\hlkwc{linetype}\hlstd{=methods,}\hlkwc{colour}\hlstd{=methods ))}\hlopt{+}
  \hlkwd{geom_point}\hlstd{(}\hlkwd{aes}\hlstd{(}\hlkwc{shape}\hlstd{=methods,}\hlkwc{colour}\hlstd{=methods))} \hlopt{+}
  \hlkwd{labs}\hlstd{(}\hlkwc{x} \hlstd{=} \hlstr{""}\hlstd{,} \hlkwc{y} \hlstd{=} \hlstr{"n = 40"}\hlstd{)} \hlopt{+}
  \hlkwd{facet_wrap}\hlstd{(} \hlopt{~} \hlstd{measures ,}\hlkwc{scales} \hlstd{=} \hlstr{"free_y"}\hlstd{,}\hlkwc{labeller} \hlstd{= label_parsed)}\hlopt{+}
  \hlkwd{scale_colour_manual}\hlstd{(}\hlkwc{values}\hlstd{=}\hlkwd{c}\hlstd{(}\hlstr{"blue"}\hlstd{,}\hlstr{"red"}\hlstd{,} \hlstr{"black"}\hlstd{))}\hlopt{+}
  \hlkwd{scale_linetype_manual}\hlstd{(}\hlkwc{values} \hlstd{=} \hlkwd{c}\hlstd{(}\hlstr{"dotted"}\hlstd{,}\hlstr{"dotted"}\hlstd{,}\hlstr{"dotted"}\hlstd{))}\hlopt{+}
  \hlkwd{scale_x_continuous}\hlstd{(}\hlstr{""}\hlstd{,}\hlkwc{labels}\hlstd{=}\hlkwd{c}\hlstd{(}\hlkwd{expression}\hlstd{(beta[}\hlnum{1}\hlstd{]),}\hlkwd{expression}\hlstd{(beta[}\hlnum{2}\hlstd{]),}
                            \hlkwd{expression}\hlstd{(beta[}\hlnum{3}\hlstd{]),}\hlkwd{expression}\hlstd{(beta[}\hlnum{4}\hlstd{]),}
                            \hlkwd{expression}\hlstd{(beta[}\hlnum{5}\hlstd{])))}\hlopt{+}
  \hlkwd{theme_minimal}\hlstd{()} \hlopt{+}
  \hlkwd{theme}\hlstd{(}\hlkwc{legend.position} \hlstd{=} \hlstr{"none"}\hlstd{,}
        \hlkwc{plot.title} \hlstd{=} \hlkwd{element_text}\hlstd{(}\hlkwc{hjust} \hlstd{=} \hlnum{0.5}\hlstd{),}
        \hlkwc{axis.text.x} \hlstd{=} \hlkwd{element_blank}\hlstd{(),}
        \hlkwc{axis.ticks.x}\hlstd{=}\hlkwd{element_blank}\hlstd{(),}
        \hlkwc{plot.margin}\hlstd{=}\hlkwd{unit}\hlstd{(}\hlkwd{c}\hlstd{(}\hlnum{0}\hlstd{,}\hlnum{0}\hlstd{,}\hlopt{-}\hlnum{0.1}\hlstd{,}\hlnum{0.03}\hlstd{),} \hlstr{"cm"}\hlstd{))}

\hlstd{pk} \hlkwb{<-} \hlkwd{ggplot}\hlstd{(}\hlkwd{subset}\hlstd{(dataggplot,(avalues}\hlopt{!=}\hlnum{1} \hlopt{&} \hlstd{coef}\hlopt{==}\hlnum{6}\hlstd{)),}\hlkwd{aes}\hlstd{(}\hlkwc{x} \hlstd{= avalues,} \hlkwc{y} \hlstd{= yvalues))} \hlopt{+}
  \hlkwd{geom_hline}\hlstd{(}\hlkwc{data}\hlstd{=hline_dat,} \hlkwd{aes}\hlstd{(}\hlkwc{yintercept}\hlstd{=threshold),} \hlkwc{col} \hlstd{=} \hlstr{"grey"}\hlstd{)} \hlopt{+}
  \hlkwd{geom_line}\hlstd{(}\hlkwd{aes}\hlstd{(}\hlkwc{linetype}\hlstd{=methods,}\hlkwc{colour}\hlstd{=methods ))}\hlopt{+}
  \hlkwd{geom_point}\hlstd{(}\hlkwd{aes}\hlstd{(}\hlkwc{shape}\hlstd{=methods,}\hlkwc{colour}\hlstd{=methods))} \hlopt{+}
  \hlkwd{labs}\hlstd{(}\hlkwc{x} \hlstd{=} \hlstr{""}\hlstd{,} \hlkwc{y} \hlstd{=} \hlstr{"n = 40"}\hlstd{)} \hlopt{+}
  \hlkwd{facet_wrap}\hlstd{(} \hlopt{~} \hlstd{measures ,}\hlkwc{scales} \hlstd{=} \hlstr{"free_y"}\hlstd{,}\hlkwc{labeller} \hlstd{= label_parsed)}\hlopt{+}
  \hlkwd{scale_colour_manual}\hlstd{(}\hlkwc{values}\hlstd{=}\hlkwd{c}\hlstd{(}\hlstr{"blue"}\hlstd{,}\hlstr{"red"}\hlstd{,} \hlstr{"black"}\hlstd{))}\hlopt{+}
  \hlkwd{scale_linetype_manual}\hlstd{(}\hlkwc{values} \hlstd{=} \hlkwd{c}\hlstd{(}\hlstr{"dotted"}\hlstd{,}\hlstr{"dotted"}\hlstd{,}\hlstr{"dotted"}\hlstd{))}\hlopt{+}
   \hlkwd{scale_x_continuous}\hlstd{()}\hlopt{+}
  \hlkwd{theme_minimal}\hlstd{()} \hlopt{+}
  \hlkwd{theme}\hlstd{(}\hlkwc{axis.ticks.x}\hlstd{=}\hlkwd{element_blank}\hlstd{(),}
        \hlkwc{axis.text.x} \hlstd{=} \hlkwd{element_blank}\hlstd{(),}
        \hlkwc{legend.position} \hlstd{=} \hlstr{"none"}\hlstd{)}

\hlcom{##n=80 ##}
\hlstd{p2_0.5} \hlkwb{<-} \hlkwd{ggplot}\hlstd{(}\hlkwd{subset}\hlstd{(dataggplot2,(avalues}\hlopt{==}\hlnum{2} \hlopt{&} \hlstd{coef}\hlopt{!=}\hlnum{6}\hlstd{)),}\hlkwd{aes}\hlstd{(}\hlkwc{x} \hlstd{= coef,} \hlkwc{y} \hlstd{= yvalues2))} \hlopt{+}
  \hlkwd{geom_hline}\hlstd{(}\hlkwc{data}\hlstd{=hline_dat,} \hlkwd{aes}\hlstd{(}\hlkwc{yintercept}\hlstd{=threshold),} \hlkwc{col} \hlstd{=} \hlstr{"grey"}\hlstd{)} \hlopt{+}
  \hlkwd{geom_line}\hlstd{(}\hlkwd{aes}\hlstd{(}\hlkwc{linetype}\hlstd{=methods,}\hlkwc{colour}\hlstd{=methods ))}\hlopt{+}
  \hlkwd{geom_point}\hlstd{(}\hlkwd{aes}\hlstd{(}\hlkwc{shape}\hlstd{=methods,}\hlkwc{colour}\hlstd{=methods))} \hlopt{+}
  \hlkwd{labs}\hlstd{(}\hlkwc{x} \hlstd{=} \hlstr{""}\hlstd{,} \hlkwc{y} \hlstd{=} \hlstr{"n = 80"}\hlstd{)} \hlopt{+}
  \hlkwd{facet_wrap}\hlstd{(} \hlopt{~} \hlstd{measures ,}\hlkwc{scales} \hlstd{=} \hlstr{"free_y"}\hlstd{,}\hlkwc{labeller} \hlstd{= label_parsed)}\hlopt{+}
  \hlkwd{scale_colour_manual}\hlstd{(}\hlkwc{values}\hlstd{=}\hlkwd{c}\hlstd{(}\hlstr{"blue"}\hlstd{,}\hlstr{"red"}\hlstd{,} \hlstr{"black"}\hlstd{))}\hlopt{+}
  \hlkwd{scale_linetype_manual}\hlstd{(}\hlkwc{values} \hlstd{=} \hlkwd{c}\hlstd{(}\hlstr{"dotted"}\hlstd{,}\hlstr{"dotted"}\hlstd{,}\hlstr{"dotted"}\hlstd{))}\hlopt{+}
  \hlkwd{scale_x_continuous}\hlstd{(}\hlstr{""}\hlstd{,}
                   \hlkwc{labels}\hlstd{=}\hlkwd{c}\hlstd{(}\hlkwd{expression}\hlstd{(beta[}\hlnum{1}\hlstd{]),}\hlkwd{expression}\hlstd{(beta[}\hlnum{2}\hlstd{]),}
                            \hlkwd{expression}\hlstd{(beta[}\hlnum{3}\hlstd{]),}\hlkwd{expression}\hlstd{(beta[}\hlnum{4}\hlstd{]),}
                            \hlkwd{expression}\hlstd{(beta[}\hlnum{5}\hlstd{])))}\hlopt{+}
  \hlkwd{theme_minimal}\hlstd{()} \hlopt{+}
  \hlkwd{theme}\hlstd{(}\hlkwc{legend.position} \hlstd{=} \hlstr{"none"}\hlstd{,}
        \hlkwc{plot.title} \hlstd{=} \hlkwd{element_text}\hlstd{(}\hlkwc{hjust} \hlstd{=} \hlnum{0.5}\hlstd{),}
        \hlkwc{plot.margin}\hlstd{=}\hlkwd{unit}\hlstd{(}\hlkwd{c}\hlstd{(}\hlopt{-}\hlnum{0.1}\hlstd{,}\hlnum{0}\hlstd{,}\hlopt{-}\hlnum{0.1}\hlstd{,}\hlnum{0.03}\hlstd{),} \hlstr{"cm"}\hlstd{))}

\hlstd{p2_0.75} \hlkwb{<-} \hlkwd{ggplot}\hlstd{(}\hlkwd{subset}\hlstd{(dataggplot2,(avalues}\hlopt{==}\hlnum{3} \hlopt{&} \hlstd{coef}\hlopt{!=}\hlnum{6}\hlstd{)),}\hlkwd{aes}\hlstd{(}\hlkwc{x} \hlstd{= coef,} \hlkwc{y} \hlstd{= yvalues2))} \hlopt{+}
  \hlkwd{geom_hline}\hlstd{(}\hlkwc{data}\hlstd{=hline_dat,} \hlkwd{aes}\hlstd{(}\hlkwc{yintercept}\hlstd{=threshold),} \hlkwc{col} \hlstd{=} \hlstr{"grey"}\hlstd{)} \hlopt{+}
  \hlkwd{geom_line}\hlstd{(}\hlkwd{aes}\hlstd{(}\hlkwc{linetype}\hlstd{=methods,}\hlkwc{colour}\hlstd{=methods ))}\hlopt{+}
  \hlkwd{geom_point}\hlstd{(}\hlkwd{aes}\hlstd{(}\hlkwc{shape}\hlstd{=methods,}\hlkwc{colour}\hlstd{=methods))} \hlopt{+}
  \hlkwd{labs}\hlstd{(}\hlkwc{x} \hlstd{=} \hlstr{""}\hlstd{,} \hlkwc{y} \hlstd{=} \hlstr{"n = 80"}\hlstd{)} \hlopt{+}
  \hlkwd{facet_wrap}\hlstd{(} \hlopt{~} \hlstd{measures ,}\hlkwc{scales} \hlstd{=} \hlstr{"free_y"}\hlstd{,}\hlkwc{labeller} \hlstd{= label_parsed)}\hlopt{+}
  \hlkwd{scale_colour_manual}\hlstd{(}\hlkwc{values}\hlstd{=}\hlkwd{c}\hlstd{(}\hlstr{"blue"}\hlstd{,}\hlstr{"red"}\hlstd{,} \hlstr{"black"}\hlstd{))}\hlopt{+}
  \hlkwd{scale_linetype_manual}\hlstd{(}\hlkwc{values} \hlstd{=} \hlkwd{c}\hlstd{(}\hlstr{"dotted"}\hlstd{,}\hlstr{"dotted"}\hlstd{,}\hlstr{"dotted"}\hlstd{))}\hlopt{+}
  \hlkwd{scale_x_continuous}\hlstd{(}\hlstr{""}\hlstd{,} \hlkwc{labels}\hlstd{=}\hlkwd{c}\hlstd{(}\hlkwd{expression}\hlstd{(beta[}\hlnum{1}\hlstd{]),}\hlkwd{expression}\hlstd{(beta[}\hlnum{2}\hlstd{]),}
                        \hlkwd{expression}\hlstd{(beta[}\hlnum{3}\hlstd{]),}\hlkwd{expression}\hlstd{(beta[}\hlnum{4}\hlstd{]),}
                        \hlkwd{expression}\hlstd{(beta[}\hlnum{5}\hlstd{])))}\hlopt{+}
  \hlkwd{theme_minimal}\hlstd{()} \hlopt{+}
  \hlkwd{theme}\hlstd{(}\hlkwc{legend.position} \hlstd{=} \hlstr{"none"}\hlstd{,}
        \hlkwc{plot.title} \hlstd{=} \hlkwd{element_text}\hlstd{(}\hlkwc{hjust} \hlstd{=} \hlnum{0.5}\hlstd{),}
        \hlkwc{plot.margin}\hlstd{=}\hlkwd{unit}\hlstd{(}\hlkwd{c}\hlstd{(}\hlopt{-}\hlnum{0.1}\hlstd{,}\hlnum{0}\hlstd{,}\hlopt{-}\hlnum{0.1}\hlstd{,}\hlnum{0.03}\hlstd{),} \hlstr{"cm"}\hlstd{))}

\hlstd{p2_1} \hlkwb{<-} \hlkwd{ggplot}\hlstd{(}\hlkwd{subset}\hlstd{(dataggplot2,(avalues}\hlopt{==}\hlnum{4} \hlopt{&} \hlstd{coef}\hlopt{!=}\hlnum{6}\hlstd{)),}\hlkwd{aes}\hlstd{(}\hlkwc{x} \hlstd{= coef,} \hlkwc{y} \hlstd{= yvalues2))} \hlopt{+}
  \hlkwd{geom_hline}\hlstd{(}\hlkwc{data}\hlstd{=hline_dat,} \hlkwd{aes}\hlstd{(}\hlkwc{yintercept}\hlstd{=threshold),} \hlkwc{col} \hlstd{=} \hlstr{"grey"}\hlstd{)} \hlopt{+}
  \hlkwd{geom_line}\hlstd{(}\hlkwd{aes}\hlstd{(}\hlkwc{linetype}\hlstd{=methods,}\hlkwc{colour}\hlstd{=methods ))}\hlopt{+}
  \hlkwd{geom_point}\hlstd{(}\hlkwd{aes}\hlstd{(}\hlkwc{shape}\hlstd{=methods,}\hlkwc{colour}\hlstd{=methods))} \hlopt{+}
  \hlkwd{labs}\hlstd{(}\hlkwc{x} \hlstd{=} \hlstr{""}\hlstd{,} \hlkwc{y} \hlstd{=} \hlstr{"n = 80"}\hlstd{)} \hlopt{+}
  \hlkwd{facet_wrap}\hlstd{(} \hlopt{~} \hlstd{measures ,}\hlkwc{scales} \hlstd{=} \hlstr{"free_y"}\hlstd{,}\hlkwc{labeller} \hlstd{= label_parsed)}\hlopt{+}
  \hlkwd{scale_colour_manual}\hlstd{(}\hlkwc{values}\hlstd{=}\hlkwd{c}\hlstd{(}\hlstr{"blue"}\hlstd{,}\hlstr{"red"}\hlstd{,} \hlstr{"black"}\hlstd{))}\hlopt{+}
  \hlkwd{scale_linetype_manual}\hlstd{(}\hlkwc{values} \hlstd{=} \hlkwd{c}\hlstd{(}\hlstr{"dotted"}\hlstd{,}\hlstr{"dotted"}\hlstd{,}\hlstr{"dotted"}\hlstd{))}\hlopt{+}
  \hlkwd{scale_x_continuous}\hlstd{(}\hlstr{""}\hlstd{,} \hlkwc{labels}\hlstd{=}\hlkwd{c}\hlstd{(}\hlkwd{expression}\hlstd{(beta[}\hlnum{1}\hlstd{]),}\hlkwd{expression}\hlstd{(beta[}\hlnum{2}\hlstd{]),}
                        \hlkwd{expression}\hlstd{(beta[}\hlnum{3}\hlstd{]),}\hlkwd{expression}\hlstd{(beta[}\hlnum{4}\hlstd{]),}
                        \hlkwd{expression}\hlstd{(beta[}\hlnum{5}\hlstd{])))}\hlopt{+}
  \hlkwd{theme_minimal}\hlstd{()} \hlopt{+}
  \hlkwd{theme}\hlstd{(}\hlkwc{legend.position} \hlstd{=} \hlstr{"none"}\hlstd{,}
        \hlkwc{plot.title} \hlstd{=} \hlkwd{element_text}\hlstd{(}\hlkwc{hjust} \hlstd{=} \hlnum{0.5}\hlstd{),}
        \hlkwc{plot.margin}\hlstd{=}\hlkwd{unit}\hlstd{(}\hlkwd{c}\hlstd{(}\hlopt{-}\hlnum{0.1}\hlstd{,}\hlnum{0}\hlstd{,}\hlopt{-}\hlnum{0.1}\hlstd{,}\hlnum{0.03}\hlstd{),} \hlstr{"cm"}\hlstd{))}

\hlstd{p2_1.5} \hlkwb{<-} \hlkwd{ggplot}\hlstd{(}\hlkwd{subset}\hlstd{(dataggplot2,(avalues}\hlopt{==}\hlnum{5} \hlopt{&} \hlstd{coef}\hlopt{!=}\hlnum{6}\hlstd{)),}\hlkwd{aes}\hlstd{(}\hlkwc{x} \hlstd{= coef,} \hlkwc{y} \hlstd{= yvalues2))} \hlopt{+}
  \hlkwd{geom_hline}\hlstd{(}\hlkwc{data}\hlstd{=hline_dat,} \hlkwd{aes}\hlstd{(}\hlkwc{yintercept}\hlstd{=threshold),} \hlkwc{col} \hlstd{=} \hlstr{"grey"}\hlstd{)} \hlopt{+}
  \hlkwd{geom_line}\hlstd{(}\hlkwd{aes}\hlstd{(}\hlkwc{linetype}\hlstd{=methods,}\hlkwc{colour}\hlstd{=methods ))}\hlopt{+}
  \hlkwd{geom_point}\hlstd{(}\hlkwd{aes}\hlstd{(}\hlkwc{shape}\hlstd{=methods,}\hlkwc{colour}\hlstd{=methods))} \hlopt{+}
  \hlkwd{labs}\hlstd{(}\hlkwc{x} \hlstd{=} \hlstr{""}\hlstd{,} \hlkwc{y} \hlstd{=} \hlstr{"n = 80"}\hlstd{)} \hlopt{+}
  \hlkwd{facet_wrap}\hlstd{(} \hlopt{~} \hlstd{measures ,}\hlkwc{scales} \hlstd{=} \hlstr{"free_y"}\hlstd{,}\hlkwc{labeller} \hlstd{= label_parsed)}\hlopt{+}
  \hlkwd{scale_colour_manual}\hlstd{(}\hlkwc{values}\hlstd{=}\hlkwd{c}\hlstd{(}\hlstr{"blue"}\hlstd{,}\hlstr{"red"}\hlstd{,} \hlstr{"black"}\hlstd{))}\hlopt{+}
  \hlkwd{scale_linetype_manual}\hlstd{(}\hlkwc{values} \hlstd{=} \hlkwd{c}\hlstd{(}\hlstr{"dotted"}\hlstd{,}\hlstr{"dotted"}\hlstd{,}\hlstr{"dotted"}\hlstd{))}\hlopt{+}
  \hlkwd{scale_x_continuous}\hlstd{(}\hlstr{""}\hlstd{,}\hlkwc{labels}\hlstd{=}\hlkwd{c}\hlstd{(}\hlkwd{expression}\hlstd{(beta[}\hlnum{1}\hlstd{]),}\hlkwd{expression}\hlstd{(beta[}\hlnum{2}\hlstd{]),}
                        \hlkwd{expression}\hlstd{(beta[}\hlnum{3}\hlstd{]),}\hlkwd{expression}\hlstd{(beta[}\hlnum{4}\hlstd{]),}
                         \hlkwd{expression}\hlstd{(beta[}\hlnum{5}\hlstd{])))}\hlopt{+}
  \hlkwd{theme_minimal}\hlstd{()} \hlopt{+}
  \hlkwd{theme}\hlstd{(}\hlkwc{legend.position} \hlstd{=} \hlstr{"none"}\hlstd{,}
        \hlkwc{plot.title} \hlstd{=} \hlkwd{element_text}\hlstd{(}\hlkwc{hjust} \hlstd{=} \hlnum{0.5}\hlstd{),}
        \hlkwc{plot.margin}\hlstd{=}\hlkwd{unit}\hlstd{(}\hlkwd{c}\hlstd{(}\hlopt{-}\hlnum{0.1}\hlstd{,}\hlnum{0}\hlstd{,}\hlopt{-}\hlnum{0.1}\hlstd{,}\hlnum{0.03}\hlstd{),} \hlstr{"cm"}\hlstd{))}

\hlstd{p2_k} \hlkwb{<-} \hlkwd{ggplot}\hlstd{(}\hlkwd{subset}\hlstd{(dataggplot2,(avalues}\hlopt{!=}\hlnum{1} \hlopt{&} \hlstd{coef}\hlopt{==}\hlnum{6}\hlstd{)),}\hlkwd{aes}\hlstd{(}\hlkwc{x} \hlstd{= avalues,} \hlkwc{y} \hlstd{= yvalues2))} \hlopt{+}
  \hlkwd{geom_hline}\hlstd{(}\hlkwc{data}\hlstd{=hline_dat,} \hlkwd{aes}\hlstd{(}\hlkwc{yintercept}\hlstd{=threshold),} \hlkwc{col} \hlstd{=} \hlstr{"grey"}\hlstd{)} \hlopt{+}
  \hlkwd{geom_line}\hlstd{(}\hlkwd{aes}\hlstd{(}\hlkwc{linetype}\hlstd{=methods,}\hlkwc{colour}\hlstd{=methods ))}\hlopt{+}
  \hlkwd{geom_point}\hlstd{(}\hlkwd{aes}\hlstd{(}\hlkwc{shape}\hlstd{=methods,}\hlkwc{colour}\hlstd{=methods))} \hlopt{+}
  \hlkwd{labs}\hlstd{(}\hlkwc{x} \hlstd{=} \hlkwd{expression}\hlstd{(kappa),} \hlkwc{y} \hlstd{=} \hlstr{" n = 80"}\hlstd{)} \hlopt{+}
  \hlkwd{facet_wrap}\hlstd{(} \hlopt{~} \hlstd{measures ,}\hlkwc{scales} \hlstd{=} \hlstr{"free_y"}\hlstd{,}\hlkwc{labeller} \hlstd{= label_parsed)}\hlopt{+}
  \hlkwd{scale_colour_manual}\hlstd{(}\hlkwc{values}\hlstd{=}\hlkwd{c}\hlstd{(}\hlstr{"blue"}\hlstd{,}\hlstr{"red"}\hlstd{,} \hlstr{"black"}\hlstd{))}\hlopt{+}
  \hlkwd{scale_linetype_manual}\hlstd{(}\hlkwc{values} \hlstd{=} \hlkwd{c}\hlstd{(}\hlstr{"dotted"}\hlstd{,}\hlstr{"dotted"}\hlstd{,}\hlstr{"dotted"}\hlstd{))}\hlopt{+}
  \hlkwd{scale_x_continuous}\hlstd{(}\hlkwc{labels}\hlstd{=}\hlkwd{c}\hlstd{(}\hlstr{"0.5"}\hlstd{,}\hlstr{"0.75"}\hlstd{,}\hlstr{"1"}\hlstd{,}\hlstr{"1.5"}\hlstd{))}\hlopt{+}
  \hlkwd{theme_minimal}\hlstd{()} \hlopt{+}
  \hlkwd{theme}\hlstd{(}\hlkwc{legend.position} \hlstd{=} \hlstr{"none"}\hlstd{)}
\end{alltt}
\end{kframe}
\end{knitrout}
Figure~2 is the result of
\begin{knitrout}\footnotesize
\definecolor{shadecolor}{rgb}{0.969, 0.969, 0.969}\color{fgcolor}\begin{kframe}
\begin{alltt}
\hlkwd{grid.arrange}\hlstd{(p0.5,p2_0.5)}
\end{alltt}
\end{kframe}\begin{figure}
\includegraphics[width=\maxwidth]{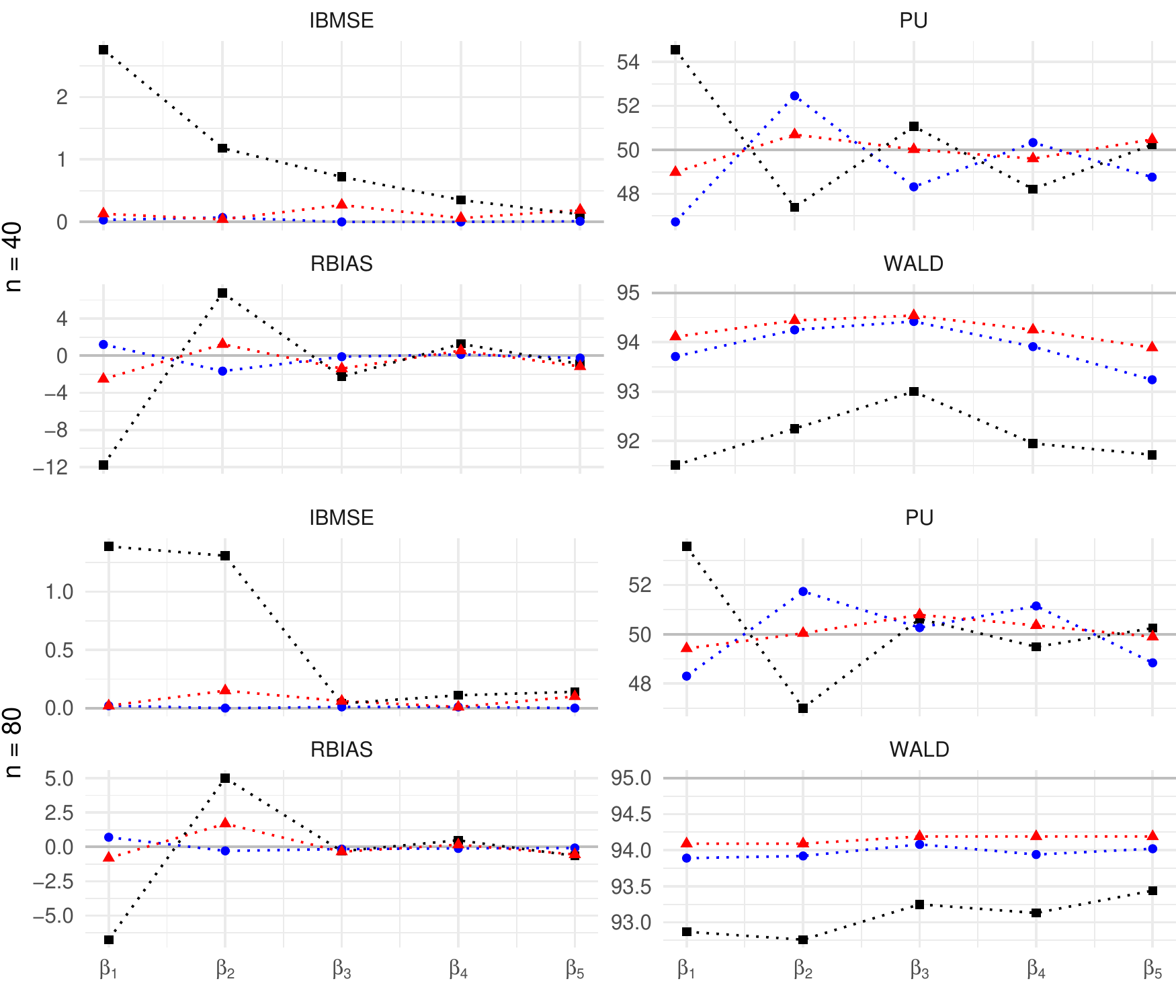} \caption[Estimation of regression parameters $\beta=(\beta_1,\beta_2,\beta_3,\beta_4,\beta_5)$ with $\kappa=0.5,n=40,80$]{Estimation of regression parameters $\beta=(\beta_1,\beta_2,\beta_3,\beta_4,\beta_5)$ with $\kappa=0.5,n=40,80$. Simulation results for ML (black squares), mean BR (blue circles) and median BR (red triangles).}\label{fig:fig2}
\end{figure}

\end{knitrout}
Figure~3 is the result of
\begin{knitrout}\footnotesize
\definecolor{shadecolor}{rgb}{0.969, 0.969, 0.969}\color{fgcolor}\begin{kframe}
\begin{alltt}
\hlkwd{grid.arrange}\hlstd{(p0.75,p2_0.75)}
\end{alltt}
\end{kframe}\begin{figure}
\includegraphics[width=\maxwidth]{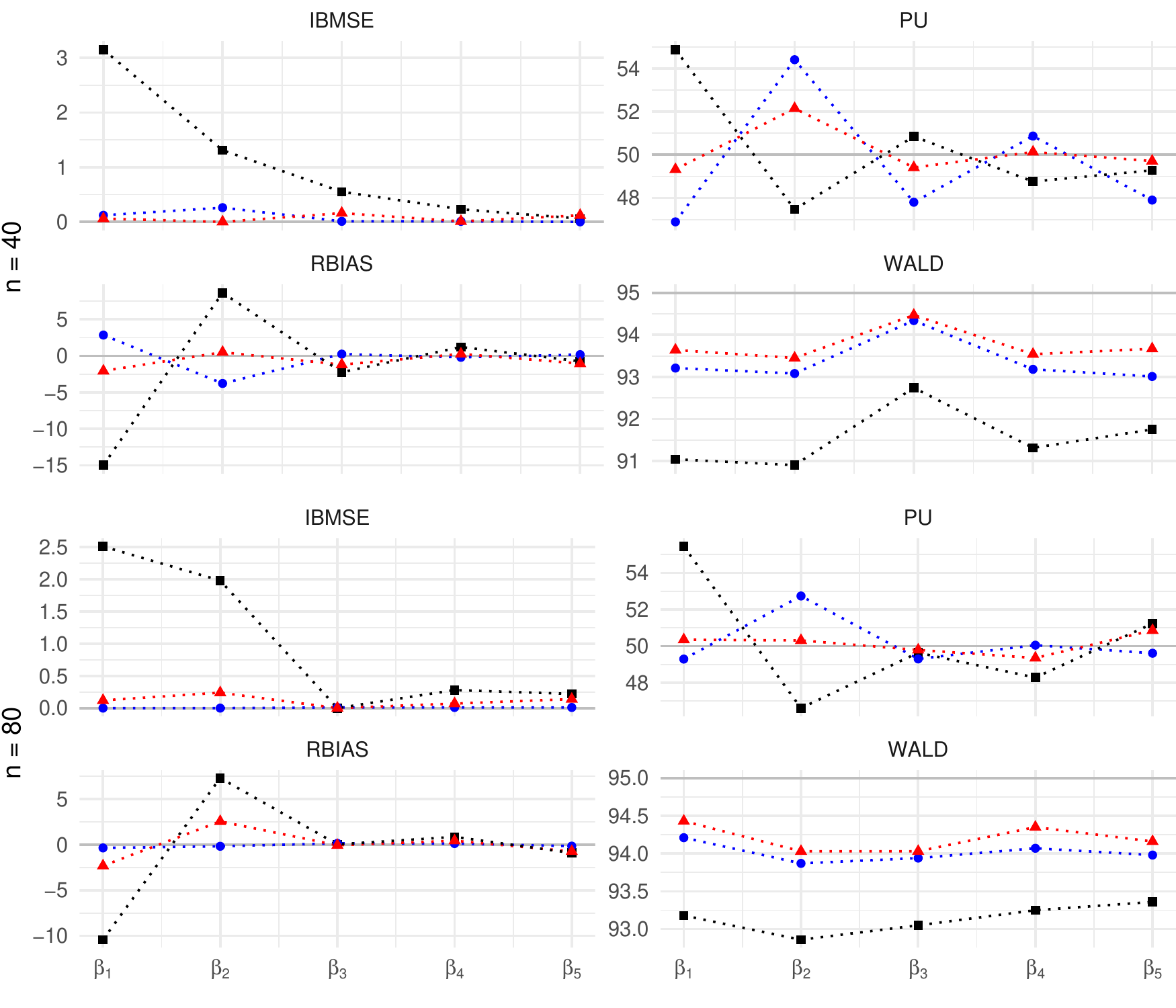} \caption[Estimation of regression parameters $\beta=(\beta_1,\beta_2,\beta_3,\beta_4,\beta_5)$ with $\kappa=0.75,n=40,80$]{Estimation of regression parameters $\beta=(\beta_1,\beta_2,\beta_3,\beta_4,\beta_5)$ with $\kappa=0.75,n=40,80$. Simulation results for ML (black squares), mean BR (blue circles) and median BR (red triangles).}\label{fig:fig3}
\end{figure}

\end{knitrout}
Figure~4 is the result of
\begin{knitrout}\footnotesize
\definecolor{shadecolor}{rgb}{0.969, 0.969, 0.969}\color{fgcolor}\begin{kframe}
\begin{alltt}
\hlkwd{grid.arrange}\hlstd{(p1,p2_1)}
\end{alltt}
\end{kframe}\begin{figure}
\includegraphics[width=\maxwidth]{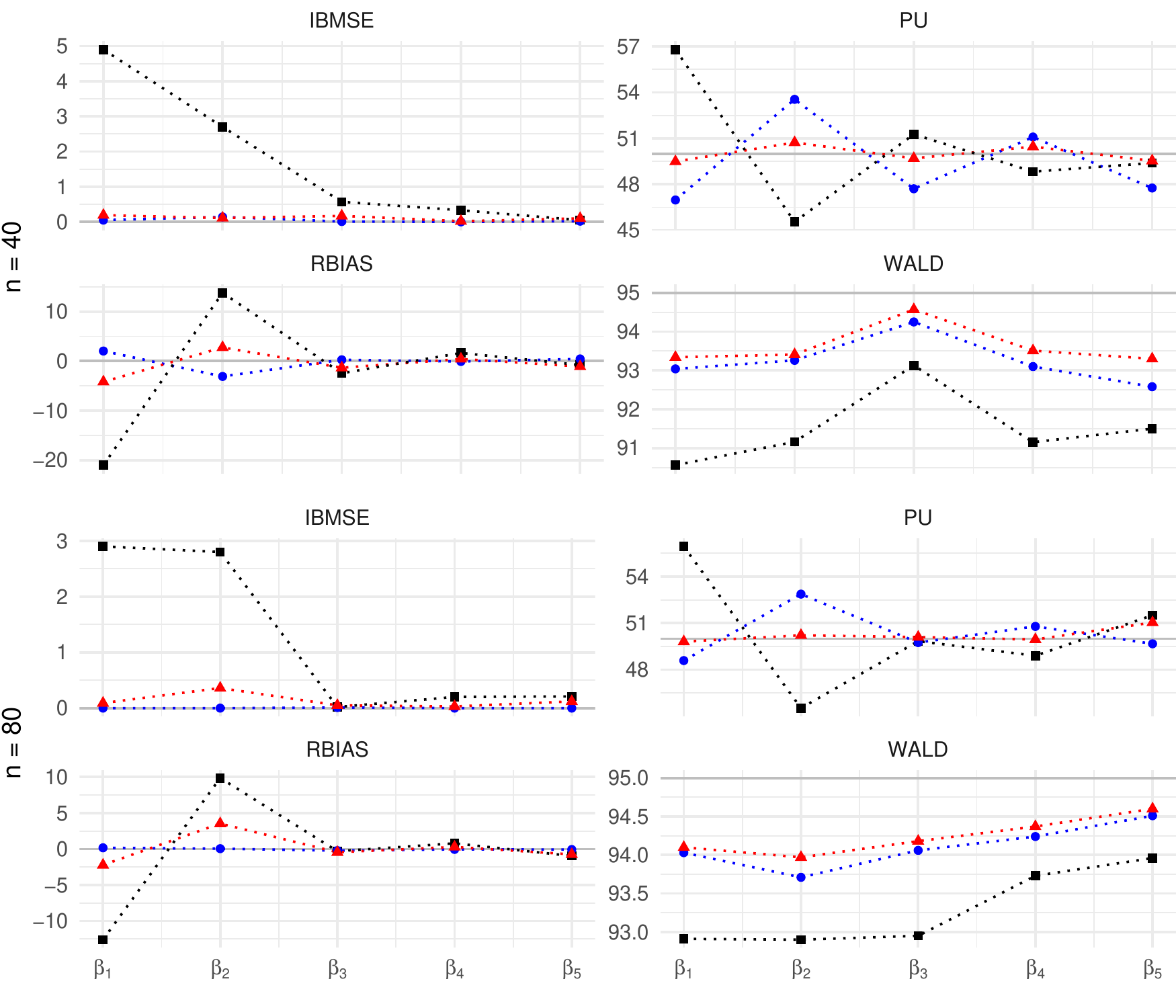} \caption[Estimation of regression parameters $\beta=(\beta_1,\beta_2,\beta_3,\beta_4,\beta_5)$ with $\kappa=1,n=40,80$]{Estimation of regression parameters $\beta=(\beta_1,\beta_2,\beta_3,\beta_4,\beta_5)$ with $\kappa=1,n=40,80$. Simulation results for ML (black squares), mean BR (blue circles) and median BR (red triangles).}\label{fig:fig4}
\end{figure}

\end{knitrout}
Figure~5 is the result of
\begin{knitrout}\footnotesize
\definecolor{shadecolor}{rgb}{0.969, 0.969, 0.969}\color{fgcolor}\begin{kframe}
\begin{alltt}
\hlkwd{grid.arrange}\hlstd{(p1.5,p2_1.5)}
\end{alltt}
\end{kframe}\begin{figure}
\includegraphics[width=\maxwidth]{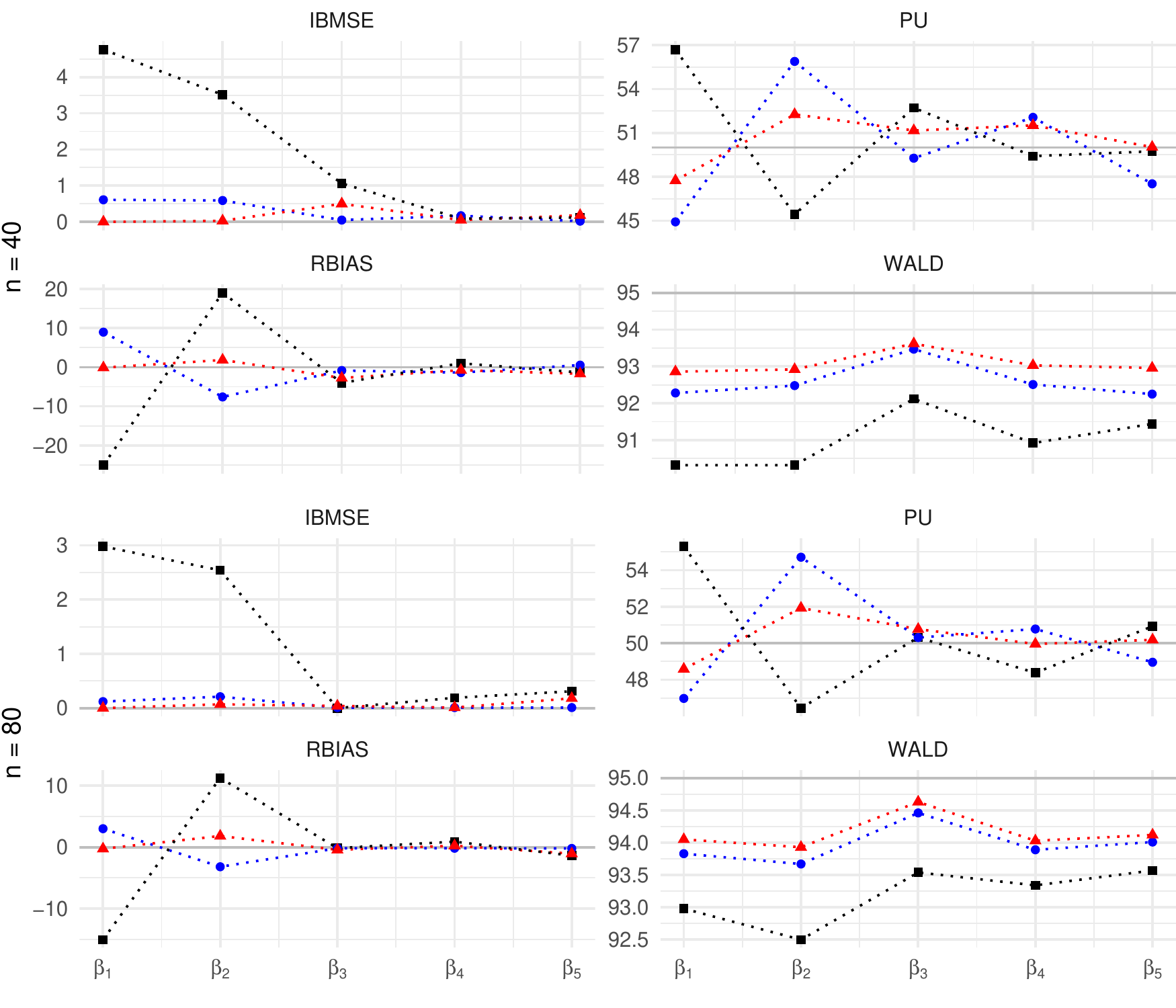} \caption[Estimation of regression parameters $\beta=(\beta_1,\beta_2,\beta_3,\beta_4,\beta_5)$ with $\kappa=1.5,n=40,80$]{Estimation of regression parameters $\beta=(\beta_1,\beta_2,\beta_3,\beta_4,\beta_5)$ with $\kappa=1.5,n=40,80$. Simulation results for ML (black squares), mean BR (blue circles) and median BR (red triangles).}\label{fig:fig5}
\end{figure}

\end{knitrout}
Figure~6 is the result of
\begin{knitrout}\footnotesize
\definecolor{shadecolor}{rgb}{0.969, 0.969, 0.969}\color{fgcolor}\begin{kframe}
\begin{alltt}
\hlkwd{grid.arrange}\hlstd{(pk,p2_k)}
\end{alltt}
\end{kframe}\begin{figure}
\includegraphics[width=\maxwidth]{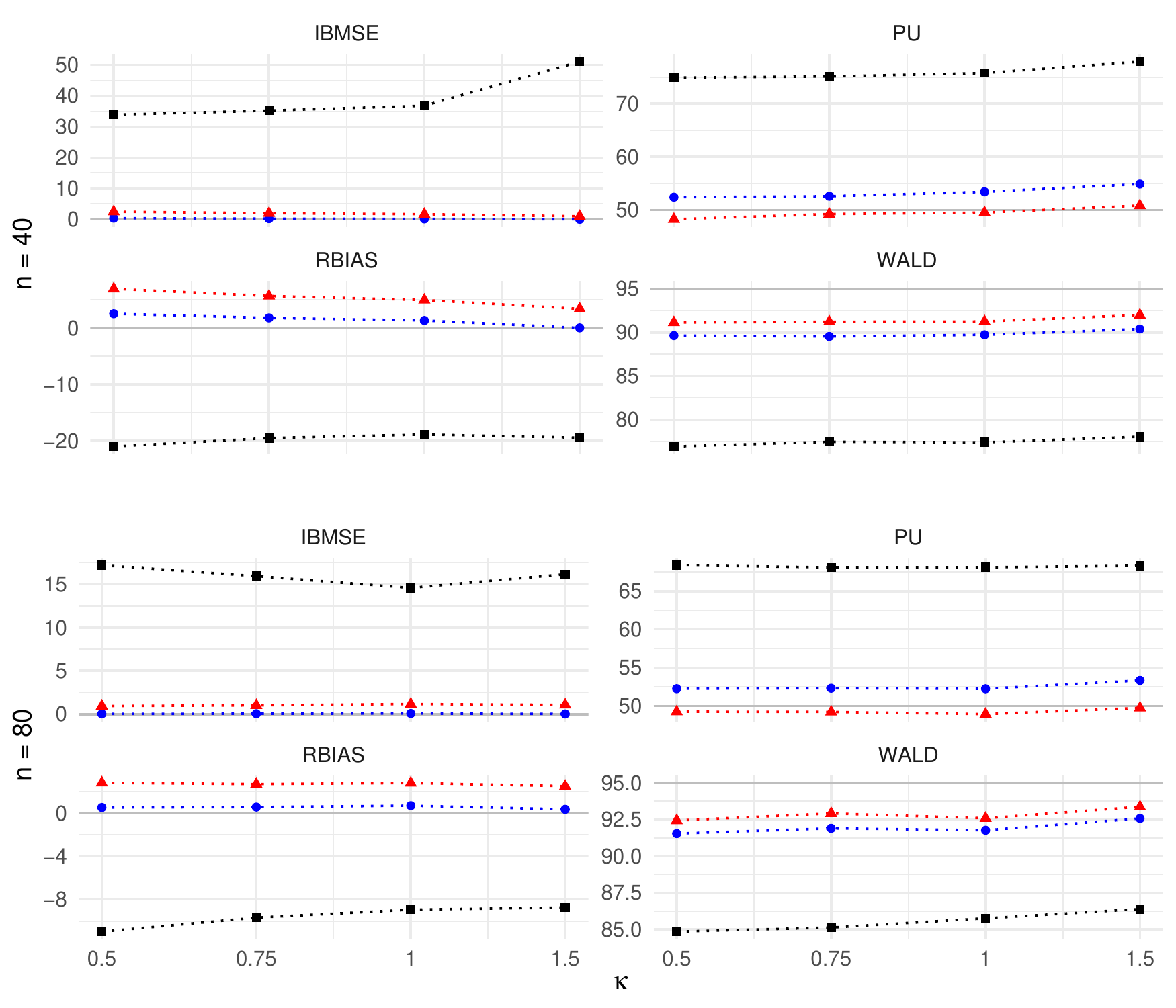} \caption[Estimation of dispersion parameter $\kappa$ with $n=40,80$]{Estimation of dispersion parameter $\kappa$ with $n=40,80$. Simulation results for ML (black squares), mean BR (blue circles) and median BR (red triangles).}\label{fig:fig6}
\end{figure}

\end{knitrout}

\section{Ames salmonella data}
This section provides the \texttt{R} code that reproduces the numerical results of Section 5.1 of the paper.
The code chunk below reproduces the results in Table 2 and illustrates the use of the \texttt{brnb} function.
\begin{knitrout}\footnotesize
\definecolor{shadecolor}{rgb}{0.969, 0.969, 0.969}\color{fgcolor}\begin{kframe}
\begin{alltt}
\hlkwd{source}\hlstd{(}\hlkwd{paste}\hlstd{(res_dir,} \hlstr{"brnb.R"}\hlstd{,} \hlkwc{sep} \hlstd{=} \hlstr{"/"}\hlstd{))}
\hlstd{freq} \hlkwb{<-} \hlkwd{c}\hlstd{(}\hlnum{15}\hlstd{,}\hlnum{16}\hlstd{,}\hlnum{16}\hlstd{,}\hlnum{27}\hlstd{,}\hlnum{33}\hlstd{,}\hlnum{20}\hlstd{,}
       \hlnum{21}\hlstd{,}\hlnum{18}\hlstd{,}\hlnum{26}\hlstd{,}\hlnum{41}\hlstd{,}\hlnum{38}\hlstd{,}\hlnum{27}\hlstd{,}
       \hlnum{29}\hlstd{,}\hlnum{21}\hlstd{,}\hlnum{33}\hlstd{,}\hlnum{60}\hlstd{,}\hlnum{41}\hlstd{,}\hlnum{42}\hlstd{)}
\hlstd{dose} \hlkwb{<-} \hlkwd{rep}\hlstd{(}\hlkwd{c}\hlstd{(}\hlnum{0}\hlstd{,}\hlnum{10}\hlstd{,}\hlnum{33}\hlstd{,}\hlnum{100}\hlstd{,}\hlnum{333}\hlstd{,}\hlnum{1000}\hlstd{),}\hlnum{3}\hlstd{)}
\hlstd{observation} \hlkwb{<-} \hlkwd{rep}\hlstd{(}\hlnum{1}\hlopt{:}\hlnum{3}\hlstd{,}\hlkwc{each}\hlstd{=}\hlnum{6}\hlstd{)}
\hlstd{salmonella} \hlkwb{<-} \hlkwd{data.frame}\hlstd{(freq,dose,observation)}

\hlstd{fitmle} \hlkwb{<-} \hlkwd{brnb}\hlstd{(freq}\hlopt{~}\hlstd{dose}\hlopt{+}\hlkwd{log}\hlstd{(dose}\hlopt{+}\hlnum{10}\hlstd{),}\hlkwc{link}\hlstd{=}\hlstr{"log"}\hlstd{,}\hlkwc{transformation} \hlstd{=}\hlstr{"identity"}\hlstd{,}
               \hlkwc{type} \hlstd{=} \hlstr{"ML"}\hlstd{,} \hlkwc{data} \hlstd{= salmonella)}
\hlstd{fitmeanBR} \hlkwb{<-} \hlkwd{brnb}\hlstd{(freq}\hlopt{~}\hlstd{dose}\hlopt{+}\hlkwd{log}\hlstd{(dose}\hlopt{+}\hlnum{10}\hlstd{),}\hlkwc{link}\hlstd{=}\hlstr{"log"}\hlstd{,}\hlkwc{transformation} \hlstd{=}\hlstr{"identity"}\hlstd{,}
              \hlkwc{type} \hlstd{=} \hlstr{"AS_mean"}\hlstd{,} \hlkwc{data} \hlstd{= salmonella)}
\hlstd{fitmedianBR} \hlkwb{<-} \hlkwd{brnb}\hlstd{(freq}\hlopt{~}\hlstd{dose}\hlopt{+}\hlkwd{log}\hlstd{(dose}\hlopt{+}\hlnum{10}\hlstd{),}\hlkwc{link}\hlstd{=}\hlstr{"log"}\hlstd{,}\hlkwc{transformation} \hlstd{=}\hlstr{"identity"}\hlstd{,}
               \hlkwc{type} \hlstd{=} \hlstr{"AS_median"}\hlstd{,} \hlkwc{data} \hlstd{= salmonella)}
\hlstd{fitmeanBC} \hlkwb{<-}  \hlkwd{brnb}\hlstd{(freq}\hlopt{~}\hlstd{dose}\hlopt{+}\hlkwd{log}\hlstd{(dose}\hlopt{+}\hlnum{10}\hlstd{),}\hlkwc{link}\hlstd{=}\hlstr{"log"}\hlstd{,}\hlkwc{transformation} \hlstd{=}\hlstr{"identity"}\hlstd{,}
                 \hlkwc{type} \hlstd{=} \hlstr{"correction"}\hlstd{,} \hlkwc{data} \hlstd{= salmonella)}
\hlstd{res} \hlkwb{<-} \hlkwd{round}\hlstd{(}\hlkwd{cbind}\hlstd{(}\hlkwd{coef}\hlstd{(fitmle,}\hlstr{"f"}\hlstd{),}\hlkwd{sqrt}\hlstd{(}\hlkwd{diag}\hlstd{(}\hlkwd{vcov}\hlstd{(fitmle,}\hlstr{"f"}\hlstd{))),}
                \hlkwd{coef}\hlstd{(fitmeanBC,}\hlstr{"f"}\hlstd{),}\hlkwd{sqrt}\hlstd{(}\hlkwd{diag}\hlstd{(}\hlkwd{vcov}\hlstd{(fitmeanBC,}\hlstr{"f"}\hlstd{))),}
                 \hlkwd{coef}\hlstd{(fitmeanBR,}\hlstr{"f"}\hlstd{),}\hlkwd{sqrt}\hlstd{(}\hlkwd{diag}\hlstd{(}\hlkwd{vcov}\hlstd{(fitmeanBR,}\hlstr{"f"}\hlstd{))),}
                \hlkwd{coef}\hlstd{(fitmedianBR,}\hlstr{"f"}\hlstd{),}\hlkwd{sqrt}\hlstd{(}\hlkwd{diag}\hlstd{(}\hlkwd{vcov}\hlstd{(fitmedianBR,}\hlstr{"f"}\hlstd{)))),}\hlnum{5}\hlstd{)}
\hlkwd{dimnames}\hlstd{(res)} \hlkwb{<-} \hlkwd{list}\hlstd{(}\hlkwd{c}\hlstd{(}\hlstr{"beta0"}\hlstd{,}\hlstr{"beta1"}\hlstd{,}\hlstr{"beta2"}\hlstd{,}\hlstr{"kappa"}\hlstd{),}\hlkwd{c}\hlstd{(}\hlstr{"mle"}\hlstd{,}\hlstr{"semle"}\hlstd{,}\hlstr{"meanBC"}\hlstd{,}\hlstr{"semeanBC"}\hlstd{,}
                                        \hlstr{"meanBR"}\hlstd{,}\hlstr{"semeanBR"}\hlstd{,}\hlstr{"medianBR"}\hlstd{,}\hlstr{"semedianBR"}\hlstd{))}
\hlstd{res}
\end{alltt}
\begin{verbatim}
##            mle   semle   meanBC semeanBC   meanBR semeanBR medianBR semedianBR
## beta0  2.19763 0.32459  2.20982  0.34817  2.21551  0.35153  2.21139    0.35918
## beta1 -0.00098 0.00039 -0.00096  0.00042 -0.00096  0.00042 -0.00096    0.00043
## beta2  0.31251 0.08790  0.31051  0.09466  0.30916  0.09563  0.30909    0.09780
## kappa  0.04877 0.02815  0.06264  0.03276  0.06473  0.03345  0.06922    0.03501
\end{verbatim}
\end{kframe}
\end{knitrout}
The following code chunk uses the image file
\texttt{salmonella\_simulation\_results.rda} to reproduce the results in Table 3  in the main text. 

\texttt{salmonella\_simulation\_results.rda}
results by running the script\\ \texttt{brnb\_salmonella\_functions.R}
which is available in the supplementary code archive.
\begin{knitrout}\footnotesize
\definecolor{shadecolor}{rgb}{0.969, 0.969, 0.969}\color{fgcolor}\begin{kframe}
\begin{alltt}
\hlkwd{load}\hlstd{(}\hlkwd{paste}\hlstd{(res_dir,} \hlstr{"salmonella_simulation_results.rda"}\hlstd{,} \hlkwc{sep} \hlstd{=} \hlstr{"/"}\hlstd{))}
\hlstd{table3}
\end{alltt}
\begin{verbatim}
##                   PU  RBIAS  WALD IBMSE
## beta0_mle      50.95  -0.62 91.77  0.17
## beta0_meanBC   49.65  -0.12 93.77  0.01
## beta0_meanBR   49.43  -0.05 93.63  0.00
## beta0_medianBR 49.98  -0.26 94.17  0.03
## beta1_mle      51.49  -1.79 91.56  0.20
## beta1_meanBC   50.21  -0.39 93.85  0.01
## beta1_meanBR   49.95  -0.09 93.65  0.00
## beta1_medianBR 50.14  -0.30 94.14  0.01
## beta2_mle      48.56   0.85 91.70  0.09
## beta2_meanBC   49.43   0.26 93.85  0.01
## beta2_meanBR   49.71   0.14 93.74  0.00
## beta2_medianBR 49.63   0.21 94.21  0.01
## kappa_mle      71.88 -22.60 81.07 20.08
## kappa_meanBC   55.37   1.98 90.73  0.11
## kappa_meanBR   53.71   3.61 89.08  0.33
## kappa_medianBR 48.44  11.96 91.56  3.37
\end{verbatim}
\end{kframe}
\end{knitrout}

\section{Epileptic seizures data}
This section provides the \texttt{R} code that reproduces the numerical results of Section 5.2 of the paper.
The code chunk below reproduces the results Figure~7 and illustrates again the use of the \texttt{brnb} function.
\begin{knitrout}\footnotesize
\definecolor{shadecolor}{rgb}{0.969, 0.969, 0.969}\color{fgcolor}\begin{kframe}
\begin{alltt}
\hlkwd{source}\hlstd{(}\hlkwd{paste}\hlstd{(res_dir,} \hlstr{"nb.r"}\hlstd{,} \hlkwc{sep} \hlstd{=} \hlstr{"/"}\hlstd{))}
\hlstd{epil2} \hlkwb{<-} \hlstd{epil[epil}\hlopt{$}\hlstd{period} \hlopt{==} \hlnum{1}\hlstd{, ]}
\hlstd{epil2[}\hlstr{"period"}\hlstd{]} \hlkwb{<-} \hlkwd{rep}\hlstd{(}\hlnum{0}\hlstd{,} \hlnum{59}\hlstd{); epil2[}\hlstr{"y"}\hlstd{]} \hlkwb{<-} \hlstd{epil2[}\hlstr{"base"}\hlstd{]; epil[}\hlstr{"time"}\hlstd{]} \hlkwb{<-} \hlnum{1}\hlstd{;}
\hlstd{epil2[}\hlstr{"time"}\hlstd{]} \hlkwb{<-} \hlnum{4}
\hlstd{epil2} \hlkwb{<-} \hlkwd{rbind}\hlstd{(epil, epil2)}
\hlstd{epil2}\hlopt{$}\hlstd{pred} \hlkwb{<-} \hlkwd{unclass}\hlstd{(epil2}\hlopt{$}\hlstd{trt)} \hlopt{*} \hlstd{(epil2}\hlopt{$}\hlstd{period} \hlopt{>} \hlnum{0}\hlstd{); epil2}\hlopt{$}\hlstd{subject} \hlkwb{<-} \hlkwd{factor}\hlstd{(epil2}\hlopt{$}\hlstd{subject)}
\hlstd{epil3} \hlkwb{<-} \hlkwd{aggregate}\hlstd{(epil2,} \hlkwd{list}\hlstd{(epil2}\hlopt{$}\hlstd{subject, epil2}\hlopt{$}\hlstd{period} \hlopt{>} \hlnum{0}\hlstd{),}
\hlkwa{function}\hlstd{(}\hlkwc{x}\hlstd{)} \hlkwa{if}\hlstd{(}\hlkwd{is.numeric}\hlstd{(x))} \hlkwd{sum}\hlstd{(x)} \hlkwa{else} \hlstd{x[}\hlnum{1}\hlstd{])}
\hlstd{epil3}\hlopt{$}\hlstd{pred} \hlkwb{<-} \hlkwd{factor}\hlstd{(epil3}\hlopt{$}\hlstd{pred,}
\hlkwc{labels} \hlstd{=} \hlkwd{c}\hlstd{(}\hlstr{"base"}\hlstd{,} \hlstr{"placebo"}\hlstd{,} \hlstr{"drug"}\hlstd{))}
\hlkwd{contrasts}\hlstd{(epil3}\hlopt{$}\hlstd{pred)} \hlkwb{<-} \hlkwd{structure}\hlstd{(}\hlkwd{contr.sdif}\hlstd{(}\hlnum{3}\hlstd{),}
\hlkwc{dimnames} \hlstd{=} \hlkwd{list}\hlstd{(}\hlkwa{NULL}\hlstd{,} \hlkwd{c}\hlstd{(}\hlstr{"placebo-base"}\hlstd{,} \hlstr{"drug-placebo"}\hlstd{)))}

\hlcom{# mle with glm.nb}
\hlstd{epil3.mle.glm.nb}  \hlkwb{<-} \hlkwd{glm.nb}\hlstd{(y} \hlopt{~ -}\hlnum{1} \hlopt{+} \hlkwd{factor}\hlstd{(subject)} \hlopt{+} \hlkwd{factor}\hlstd{(pred),}  \hlkwc{data} \hlstd{= epil3)}
\hlcom{# mle}
\hlstd{epil3.mle} \hlkwb{<-} \hlkwd{brnb}\hlstd{(y} \hlopt{~ -}\hlnum{1}\hlopt{+} \hlkwd{factor}\hlstd{(subject)} \hlopt{+} \hlkwd{factor}\hlstd{(pred),}  \hlkwc{data} \hlstd{= epil3,}\hlkwc{type}\hlstd{=}\hlstr{"ML"}\hlstd{)}
\hlcom{# meanBR}
\hlstd{epil3.br} \hlkwb{<-} \hlkwd{brnb}\hlstd{(y} \hlopt{~ -}\hlnum{1}\hlopt{+} \hlkwd{factor}\hlstd{(subject)} \hlopt{+} \hlkwd{factor}\hlstd{(pred),}  \hlkwc{data} \hlstd{= epil3,}\hlkwc{type}\hlstd{=}\hlstr{"AS_mean"}\hlstd{,}
                   \hlkwc{start} \hlstd{=} \hlkwd{coef}\hlstd{(epil3.mle,}\hlstr{"full"} \hlstd{))}
\hlcom{# medianBR}
\hlstd{epil3.mbr} \hlkwb{<-} \hlkwd{brnb}\hlstd{(y} \hlopt{~ -}\hlnum{1}\hlopt{+} \hlkwd{factor}\hlstd{(subject)} \hlopt{+} \hlkwd{factor}\hlstd{(pred),}  \hlkwc{data} \hlstd{= epil3,}\hlkwc{type}\hlstd{=}\hlstr{"AS_median"}\hlstd{,}
                    \hlkwc{start} \hlstd{=} \hlkwd{coef}\hlstd{(epil3.mle,}\hlstr{"full"} \hlstd{))}
\hlcom{# meanBR }
\hlstd{epil3.bc} \hlkwb{<-} \hlkwd{brnb}\hlstd{(y} \hlopt{~ -}\hlnum{1}\hlopt{+} \hlkwd{factor}\hlstd{(subject)} \hlopt{+} \hlkwd{factor}\hlstd{(pred),}  \hlkwc{data} \hlstd{= epil3,}\hlkwc{type}\hlstd{=}\hlstr{"correction"}\hlstd{,}
                   \hlkwc{start} \hlstd{=}  \hlkwd{coef}\hlstd{(epil3.mle,}\hlstr{"full"} \hlstd{))}
\hlcom{# modified profile likelihood}
\hlstd{epil3.mpl} \hlkwb{<-} \hlkwd{nb.MPL}\hlstd{(y} \hlopt{~ -}\hlnum{1} \hlopt{+} \hlstd{pred}\hlopt{+}\hlkwd{strata}\hlstd{(subject),} \hlkwc{strata} \hlstd{= epil3}\hlopt{$}\hlstd{subject,} \hlkwc{data} \hlstd{= epil3,}
                    \hlkwc{obj.mle} \hlstd{= epil3.mle.glm.nb ,} \hlkwc{hessian} \hlstd{=} \hlnum{TRUE}\hlstd{)}

\hlstd{ml.est} \hlkwb{<-} \hlkwd{coef}\hlstd{(epil3.mle,}\hlstr{"full"}\hlstd{)[}\hlopt{-}\hlkwd{c}\hlstd{(}\hlnum{1}\hlopt{:}\hlnum{59}\hlstd{)]}
\hlstd{ml.se} \hlkwb{<-} \hlkwd{sqrt}\hlstd{(}\hlkwd{diag}\hlstd{(}\hlkwd{vcov}\hlstd{(epil3.mle,}\hlstr{"full"}\hlstd{)))[}\hlopt{-}\hlkwd{c}\hlstd{(}\hlnum{1}\hlopt{:}\hlnum{59}\hlstd{)]}
\hlstd{ml.ciU} \hlkwb{<-} \hlstd{ml.est}\hlopt{+}\hlkwd{qnorm}\hlstd{(}\hlnum{0.975}\hlstd{)}\hlopt{*}\hlstd{ml.se}
\hlstd{ml.ciL} \hlkwb{<-} \hlstd{ml.est}\hlopt{-}\hlkwd{qnorm}\hlstd{(}\hlnum{0.975}\hlstd{)}\hlopt{*}\hlstd{ml.se}

\hlstd{bc.est} \hlkwb{<-} \hlkwd{coef}\hlstd{(epil3.bc,}\hlstr{"full"}\hlstd{)[}\hlopt{-}\hlkwd{c}\hlstd{(}\hlnum{1}\hlopt{:}\hlnum{59}\hlstd{)]}
\hlstd{bc.se} \hlkwb{<-}\hlkwd{sqrt}\hlstd{(}\hlkwd{diag}\hlstd{(}\hlkwd{vcov}\hlstd{(epil3.bc,}\hlstr{"full"}\hlstd{)))[}\hlopt{-}\hlkwd{c}\hlstd{(}\hlnum{1}\hlopt{:}\hlnum{59}\hlstd{)]}
\hlstd{bc.ciU} \hlkwb{<-} \hlstd{bc.est}\hlopt{+}\hlkwd{qnorm}\hlstd{(}\hlnum{0.975}\hlstd{)}\hlopt{*}\hlstd{bc.se}
\hlstd{bc.ciL} \hlkwb{<-} \hlstd{bc.est}\hlopt{-}\hlkwd{qnorm}\hlstd{(}\hlnum{0.975}\hlstd{)}\hlopt{*}\hlstd{bc.se}

\hlstd{br.est} \hlkwb{<-} \hlkwd{coef}\hlstd{(epil3.br,}\hlstr{"full"}\hlstd{)[}\hlopt{-}\hlkwd{c}\hlstd{(}\hlnum{1}\hlopt{:}\hlnum{59}\hlstd{)]}
\hlstd{br.se} \hlkwb{<-} \hlkwd{sqrt}\hlstd{(}\hlkwd{diag}\hlstd{(}\hlkwd{vcov}\hlstd{(epil3.br,}\hlstr{"full"}\hlstd{)))[}\hlopt{-}\hlkwd{c}\hlstd{(}\hlnum{1}\hlopt{:}\hlnum{59}\hlstd{)]}
\hlstd{br.ciU} \hlkwb{<-} \hlstd{br.est}\hlopt{+}\hlkwd{qnorm}\hlstd{(}\hlnum{0.975}\hlstd{)}\hlopt{*}\hlstd{br.se}
\hlstd{br.ciL} \hlkwb{<-} \hlstd{br.est}\hlopt{-}\hlkwd{qnorm}\hlstd{(}\hlnum{0.975}\hlstd{)}\hlopt{*}\hlstd{br.se}

\hlstd{mbr.est} \hlkwb{<-} \hlkwd{coef}\hlstd{(epil3.mbr,}\hlstr{"full"}\hlstd{)[}\hlopt{-}\hlkwd{c}\hlstd{(}\hlnum{1}\hlopt{:}\hlnum{59}\hlstd{)]}
\hlstd{mbr.se} \hlkwb{<-} \hlkwd{sqrt}\hlstd{(}\hlkwd{diag}\hlstd{(}\hlkwd{vcov}\hlstd{(epil3.mbr,}\hlstr{"full"}\hlstd{)))[}\hlopt{-}\hlkwd{c}\hlstd{(}\hlnum{1}\hlopt{:}\hlnum{59}\hlstd{)]}
\hlstd{mbr.ciU} \hlkwb{<-} \hlstd{mbr.est}\hlopt{+}\hlkwd{qnorm}\hlstd{(}\hlnum{0.975}\hlstd{)}\hlopt{*}\hlstd{mbr.se}
\hlstd{mbr.ciL} \hlkwb{<-} \hlstd{mbr.est}\hlopt{-}\hlkwd{qnorm}\hlstd{(}\hlnum{0.975}\hlstd{)}\hlopt{*}\hlstd{mbr.se}

\hlstd{mpl.est} \hlkwb{<-} \hlstd{epil3.mpl}\hlopt{$}\hlstd{coef}
\hlstd{mpl.est[}\hlnum{3}\hlstd{]} \hlkwb{<-} \hlkwd{exp}\hlstd{(}\hlopt{-}\hlstd{epil3.mpl}\hlopt{$}\hlstd{coef[}\hlnum{3}\hlstd{])}
\hlstd{mpl.se} \hlkwb{<-}  \hlstd{epil3.mpl}\hlopt{$}\hlstd{se}
\hlstd{mpl.se[}\hlnum{3}\hlstd{]} \hlkwb{<-} \hlstd{mpl.se[}\hlnum{3}\hlstd{]}\hlopt{*}\hlkwd{exp}\hlstd{(}\hlopt{-}\hlstd{epil3.mpl}\hlopt{$}\hlstd{coef[}\hlnum{3}\hlstd{])}
\hlstd{mpl.ciU} \hlkwb{<-} \hlstd{mpl.est}\hlopt{+}\hlkwd{qnorm}\hlstd{(}\hlnum{0.975}\hlstd{)}\hlopt{*}\hlstd{mpl.se}
\hlstd{mpl.ciL} \hlkwb{<-} \hlstd{mpl.est}\hlopt{-}\hlkwd{qnorm}\hlstd{(}\hlnum{0.975}\hlstd{)}\hlopt{*}\hlstd{mpl.se}
\end{alltt}
\end{kframe}
\end{knitrout}
Figure~7 is the results of
\begin{knitrout}\footnotesize
\definecolor{shadecolor}{rgb}{0.969, 0.969, 0.969}\color{fgcolor}\begin{kframe}
\begin{alltt}
\hlkwd{plot}\hlstd{(}\hlnum{1}\hlopt{:}\hlnum{3}\hlstd{,ml.ciU,}\hlkwc{type}\hlstd{=}\hlstr{"n"}\hlstd{,}\hlkwc{ylim}\hlstd{=}\hlkwd{range}\hlstd{(bc.ciL,bc.ciU,ml.ciU,ml.ciL,mpl.ciU,mpl.ciL,}
                                    \hlstd{br.ciU,br.ciL,mbr.ciU,mbr.ciL),}
     \hlkwc{xlim}\hlstd{=}\hlkwd{c}\hlstd{(}\hlnum{0.5}\hlstd{,}\hlnum{4}\hlstd{),}\hlkwc{ylab}\hlstd{=}\hlstr{"95% Wald confidence interval"}\hlstd{,}\hlkwc{xaxt}\hlstd{=}\hlstr{"n"}\hlstd{,}\hlkwc{xlab}\hlstd{=}\hlstr{"Parameters"}\hlstd{)}
\hlkwd{axis}\hlstd{(}\hlnum{1}\hlstd{,}\hlkwc{at}\hlstd{=}\hlnum{1}\hlopt{:}\hlnum{3}\hlstd{,}\hlkwc{labels}\hlstd{=}\hlkwd{expression}\hlstd{(beta[}\hlnum{1}\hlstd{],beta[}\hlnum{2}\hlstd{],kappa))}
\hlkwd{segments}\hlstd{((}\hlnum{1}\hlopt{:}\hlnum{3}\hlstd{),mpl.ciU,(}\hlnum{1}\hlopt{:}\hlnum{3}\hlstd{),mpl.ciL,}\hlkwc{col}\hlstd{=}\hlstr{"gold"}\hlstd{,}\hlkwc{lwd}\hlstd{=}\hlnum{2}\hlstd{,}\hlkwc{lty}\hlstd{=}\hlstr{"longdash"}\hlstd{)}
\hlkwd{points}\hlstd{((}\hlnum{1}\hlopt{:}\hlnum{3}\hlstd{),mpl.est,}\hlkwc{col}\hlstd{=}\hlstr{"gold"}\hlstd{,}\hlkwc{pch}\hlstd{=}\hlnum{20}\hlstd{)}
\hlkwd{segments}\hlstd{((}\hlnum{1}\hlopt{:}\hlnum{3}\hlstd{)}\hlopt{+}\hlnum{0.1}\hlstd{,bc.ciU,(}\hlnum{1}\hlopt{:}\hlnum{3}\hlstd{)}\hlopt{+}\hlnum{0.1}\hlstd{,bc.ciL,}\hlkwc{col}\hlstd{=}\hlstr{"green"}\hlstd{,}\hlkwc{lwd}\hlstd{=}\hlnum{2}\hlstd{,}\hlkwc{lty}\hlstd{=}\hlstr{"dotted"}\hlstd{)}
\hlkwd{points}\hlstd{((}\hlnum{1}\hlopt{:}\hlnum{3}\hlstd{)}\hlopt{+}\hlnum{0.1}\hlstd{,bc.est,}\hlkwc{col}\hlstd{=}\hlstr{"green"}\hlstd{,}\hlkwc{pch}\hlstd{=}\hlnum{20}\hlstd{)}
\hlkwd{segments}\hlstd{((}\hlnum{1}\hlopt{:}\hlnum{3}\hlstd{)}\hlopt{+}\hlnum{0.20}\hlstd{,ml.ciU,(}\hlnum{1}\hlopt{:}\hlnum{3}\hlstd{)}\hlopt{+}\hlnum{0.20}\hlstd{,ml.ciL,}\hlkwc{col}\hlstd{=}\hlnum{1}\hlstd{,}\hlkwc{lwd}\hlstd{=}\hlnum{2}\hlstd{,} \hlkwc{lty}\hlstd{=}\hlnum{1}\hlstd{)}
\hlkwd{points}\hlstd{((}\hlnum{1}\hlopt{:}\hlnum{3}\hlstd{)}\hlopt{+}\hlnum{0.20}\hlstd{,ml.est,}\hlkwc{col}\hlstd{=}\hlnum{1}\hlstd{,}\hlkwc{pch}\hlstd{=}\hlnum{20}\hlstd{)}
\hlkwd{segments}\hlstd{((}\hlnum{1}\hlopt{:}\hlnum{3}\hlstd{)}\hlopt{-}\hlnum{0.1}\hlstd{,br.ciU,(}\hlnum{1}\hlopt{:}\hlnum{3}\hlstd{)}\hlopt{-}\hlnum{0.1}\hlstd{,br.ciL,}\hlkwc{col}\hlstd{=}\hlstr{"red"}\hlstd{,}\hlkwc{lwd}\hlstd{=}\hlnum{2}\hlstd{,}\hlkwc{lty}\hlstd{=}\hlstr{"dashed"}\hlstd{)}
\hlkwd{points}\hlstd{((}\hlnum{1}\hlopt{:}\hlnum{3}\hlstd{)}\hlopt{-}\hlnum{0.1}\hlstd{,br.est,}\hlkwc{col}\hlstd{=}\hlstr{"red"}\hlstd{,}\hlkwc{pch}\hlstd{=}\hlnum{20}\hlstd{)}
\hlkwd{segments}\hlstd{((}\hlnum{1}\hlopt{:}\hlnum{3}\hlstd{)}\hlopt{-}\hlnum{0.2}\hlstd{,mbr.ciU,(}\hlnum{1}\hlopt{:}\hlnum{3}\hlstd{)}\hlopt{-}\hlnum{0.2}\hlstd{,mbr.ciL,}\hlkwc{col}\hlstd{=}\hlstr{"blue"}\hlstd{,}\hlkwc{lwd}\hlstd{=}\hlnum{2}\hlstd{,}\hlkwc{lty}\hlstd{=}\hlstr{"dotdash"}\hlstd{)}
\hlkwd{points}\hlstd{((}\hlnum{1}\hlopt{:}\hlnum{3}\hlstd{)}\hlopt{-}\hlnum{0.2}\hlstd{,mbr.est,}\hlkwc{col}\hlstd{=}\hlstr{"blue"}\hlstd{,}\hlkwc{pch}\hlstd{=}\hlnum{20}\hlstd{)}
\hlkwd{abline}\hlstd{(}\hlkwc{v}\hlstd{=}\hlkwd{c}\hlstd{(}\hlnum{0.5}\hlstd{,}\hlnum{1.5}\hlstd{,}\hlnum{2.5}\hlstd{,}\hlnum{3.5}\hlstd{),}\hlkwc{col}\hlstd{=}\hlstr{"gray"}\hlstd{,}\hlkwc{lty}\hlstd{=}\hlstr{"dotted"}\hlstd{)}
\hlkwd{legend}\hlstd{(}\hlkwc{legend}\hlstd{=}\hlkwd{c}\hlstd{(}\hlstr{"MPL"}\hlstd{,}\hlstr{"mean BC"}\hlstd{,}\hlstr{"ML"}\hlstd{,}\hlstr{"mean BR"}\hlstd{,}\hlstr{"median BR"}\hlstd{),}\hlkwc{x}\hlstd{=}\hlnum{2.8}\hlstd{,}\hlkwc{y}\hlstd{=}\hlopt{-}\hlnum{0.2}\hlstd{,}
       \hlkwc{col}\hlstd{=}\hlkwd{c}\hlstd{(}\hlstr{"gold"}\hlstd{,}\hlstr{"green"}\hlstd{,}\hlstr{"black"}\hlstd{,}\hlstr{"red"}\hlstd{,}\hlstr{"blue"}\hlstd{),}
       \hlkwc{lty}\hlstd{=}\hlkwd{c}\hlstd{(}\hlstr{"longdash"}\hlstd{,}\hlstr{"dotted"}\hlstd{,}\hlstr{"solid"}\hlstd{,}\hlstr{"dashed"}\hlstd{,}\hlstr{"dotdash"}\hlstd{),}\hlkwc{lwd}\hlstd{=}\hlnum{2}\hlstd{,}\hlkwc{bty} \hlstd{=} \hlstr{"n"}\hlstd{)}
\end{alltt}
\end{kframe}\begin{figure}
\includegraphics[width=\maxwidth]{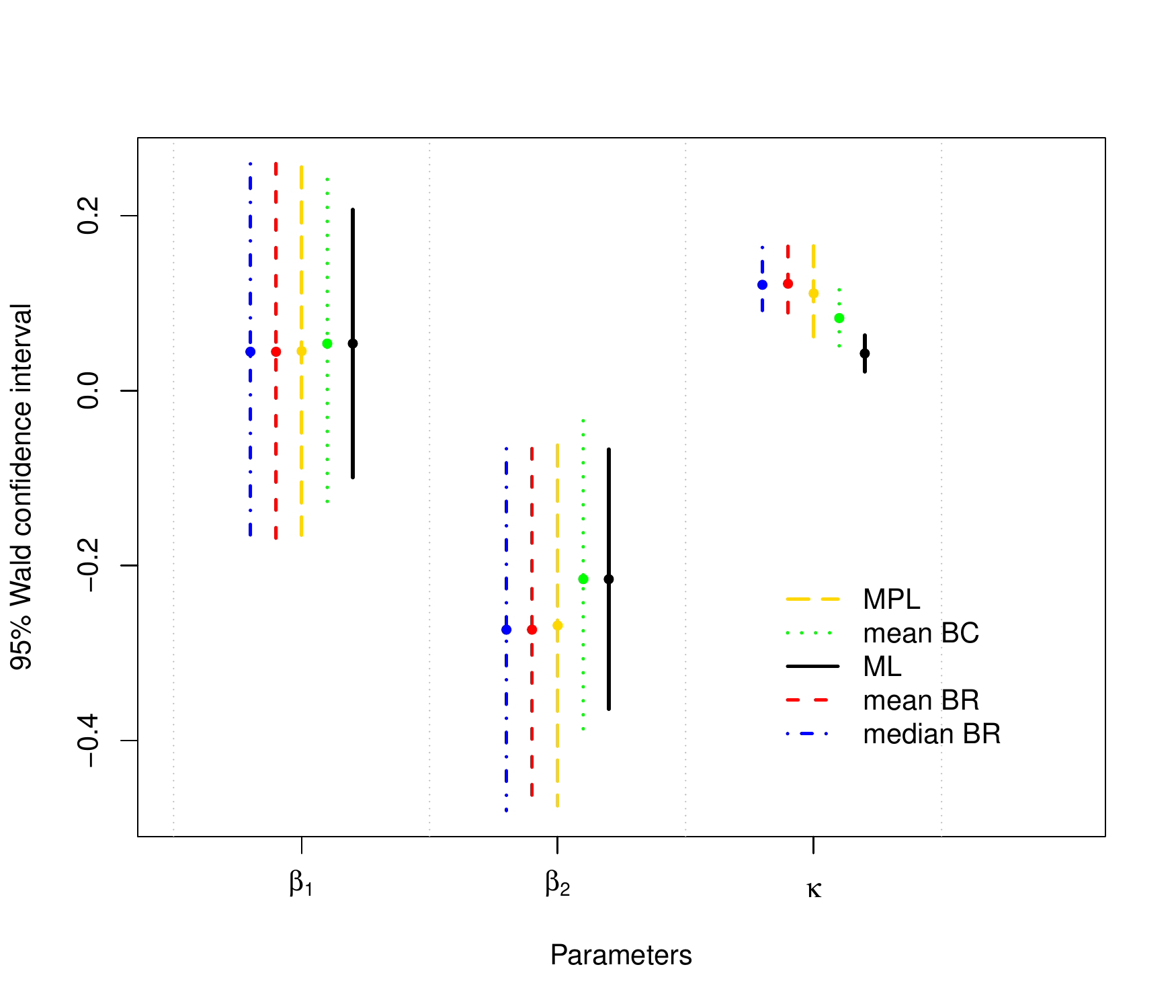} \caption[Epileptic seizures]{Epileptic seizures: points represent the parameter estimates while the vertical lines represent 95\% Wald-type confidence intervals.}\label{fig:fig7b}
\end{figure}

\end{knitrout}

The following code chunk uses the image file
\texttt{epileptic\_simulation\_results.rda} to reproduce the results in Table 4 and Figure 8  in the main text. 

\texttt{epileptic\_simulation\_results.rda}
results by running the script\\ \texttt{brnb\_epileptic\_functions.R}
which is available in the supplementary code archive. Table~4 is the result of 
\begin{knitrout}\footnotesize
\definecolor{shadecolor}{rgb}{0.969, 0.969, 0.969}\color{fgcolor}\begin{kframe}
\begin{alltt}
\hlkwd{load}\hlstd{(}\hlkwd{paste}\hlstd{(res_dir,} \hlstr{"epileptic_simulation_results.rda"}\hlstd{,} \hlkwc{sep} \hlstd{=} \hlstr{"/"}\hlstd{))}
\hlstd{table4}
\end{alltt}
\begin{verbatim}
##                    PU  RBIAS  WALD   IBMSE
## beta1_mle       49.80   0.42 82.22    0.00
## beta1_meanBC    50.15  -0.16 89.63    0.00
## beta1_meanBR    49.86   0.52 94.36    0.00
## beta1_medianBR  49.89   0.54 94.40    0.00
## beta2_mle       49.84   0.09 82.03    0.00
## beta2_meanBC    49.45   0.37 89.04    0.01
## beta2_meanBR    50.63  -0.50 94.55    0.02
## beta2_medianBR  50.62  -0.53 94.60    0.02
## kappa_mle      100.00 -79.32  0.39 3671.91
## kappa_meanBC    93.47 -40.99 40.34  286.52
## kappa_meanBR    48.78   3.99 81.44    1.10
## kappa_medianBR  48.81   3.89 82.15    1.08
\end{verbatim}
\end{kframe}
\end{knitrout}

The code chunk below prepares the data  for producing Figure~8 in the main text..
\begin{knitrout}\footnotesize
\definecolor{shadecolor}{rgb}{0.969, 0.969, 0.969}\color{fgcolor}\begin{kframe}
\begin{alltt}
\hlstd{yvalues}\hlkwb{=}\hlkwd{c}\hlstd{(rbias_nuis,pu_nuis,cov_nuis,ibmse_nuis)}
\hlstd{methods}\hlkwb{=}\hlkwd{rep}\hlstd{(}\hlkwd{rep}\hlstd{(}\hlkwd{c}\hlstd{(}\hlstr{"1"}\hlstd{,}\hlstr{"2"}\hlstd{,}\hlstr{"3"}\hlstd{,}\hlstr{"4"}\hlstd{),}\hlkwc{each}\hlstd{=}\hlnum{59}\hlstd{),}\hlnum{4}\hlstd{)}
\hlstd{measures}\hlkwb{=} \hlkwd{as.factor}\hlstd{(}\hlkwd{rep}\hlstd{(}\hlkwd{c}\hlstd{(}\hlnum{1}\hlstd{,}\hlnum{2}\hlstd{,}\hlnum{3}\hlstd{,}\hlnum{4}\hlstd{),}\hlkwc{each}\hlstd{=}\hlnum{236}\hlstd{))}
\hlkwd{levels}\hlstd{(measures)}\hlkwb{=}\hlkwd{c}\hlstd{(}\hlstr{"RBIAS"}\hlstd{,}\hlstr{"PU"}\hlstd{,}\hlstr{"WALD"}\hlstd{,}\hlstr{"IBMSE"}\hlstd{)}
\hlstd{xvalues} \hlkwb{=} \hlkwd{rep}\hlstd{(}\hlkwd{rep}\hlstd{(}\hlnum{1}\hlopt{:}\hlnum{59}\hlstd{,}\hlnum{4}\hlstd{),}\hlnum{4}\hlstd{)}
\hlstd{dataggplot} \hlkwb{=} \hlkwd{data.frame}\hlstd{(yvalues,methods,xvalues,measures)}
\hlstd{hline_dat}\hlkwb{=}\hlkwd{data.frame}\hlstd{(}\hlkwc{measures}\hlstd{=}\hlkwd{c}\hlstd{(}\hlstr{"RBIAS"} \hlstd{,}  \hlstr{"PU"}\hlstd{,}   \hlstr{"WALD"}\hlstd{,} \hlstr{"IBMSE"}\hlstd{),}
                     \hlkwc{threshold}\hlstd{=}\hlkwd{c}\hlstd{(}\hlnum{0}\hlstd{,} \hlnum{50}\hlstd{,} \hlnum{95}\hlstd{,} \hlnum{0}\hlstd{))}
\end{alltt}
\end{kframe}
\end{knitrout}

Figure~8 is the results of
\begin{knitrout}\footnotesize
\definecolor{shadecolor}{rgb}{0.969, 0.969, 0.969}\color{fgcolor}\begin{kframe}
\begin{alltt}
\hlkwd{ggplot}\hlstd{(dataggplot,}\hlkwd{aes}\hlstd{(}\hlkwc{x} \hlstd{= xvalues,} \hlkwc{y} \hlstd{= yvalues))} \hlopt{+}
  \hlkwd{geom_hline}\hlstd{(}\hlkwc{data}\hlstd{=hline_dat,} \hlkwd{aes}\hlstd{(}\hlkwc{yintercept}\hlstd{=threshold),} \hlkwc{col} \hlstd{=} \hlstr{"grey"}\hlstd{)} \hlopt{+}
  \hlkwd{geom_line}\hlstd{(}\hlkwd{aes}\hlstd{(}\hlkwc{linetype}\hlstd{=methods,}\hlkwc{colour}\hlstd{=methods ))}\hlopt{+}
  \hlkwd{ggtitle}\hlstd{(}\hlstr{""}\hlstd{)} \hlopt{+}
  \hlkwd{labs}\hlstd{(}\hlkwc{x} \hlstd{=} \hlstr{""}\hlstd{,} \hlkwc{y} \hlstd{=} \hlstr{""}\hlstd{)} \hlopt{+}
  \hlkwd{facet_wrap}\hlstd{(} \hlopt{~}\hlstd{measures  ,}\hlkwc{scales} \hlstd{=} \hlstr{"free_y"}\hlstd{,}\hlkwc{labeller} \hlstd{= label_parsed,}\hlkwc{nrow}\hlstd{=}\hlnum{4}\hlstd{)}\hlopt{+}
  \hlkwd{scale_colour_manual}\hlstd{(}\hlkwc{values}\hlstd{=}\hlkwd{c}\hlstd{(}\hlstr{"black"}\hlstd{,}\hlstr{"green"}\hlstd{,} \hlstr{"red"}\hlstd{,}\hlstr{"blue"}\hlstd{))}\hlopt{+}
  \hlkwd{scale_linetype_manual}\hlstd{(}\hlkwc{values} \hlstd{=} \hlkwd{c}\hlstd{(}\hlstr{"solid"}\hlstd{,}\hlstr{"dotted"}\hlstd{,}\hlstr{"dashed"}\hlstd{,}\hlstr{"dotdash"}\hlstd{))}\hlopt{+}
  \hlkwd{scale_x_discrete}\hlstd{(}\hlkwd{expression}\hlstd{(}\hlkwd{paste}\hlstd{(}\hlstr{"Nuisance parameters"}\hlstd{,}\hlstr{" ("}\hlstd{,lambda,}\hlstr{")"}\hlstd{)),}
                   \hlkwc{limits}\hlstd{=}\hlkwd{c}\hlstd{(}\hlnum{1}\hlopt{:}\hlnum{59}\hlstd{),} \hlkwc{breaks}\hlstd{=}\hlkwd{c}\hlstd{(}\hlnum{1}\hlopt{:}\hlnum{59}\hlstd{),}\hlkwc{labels}\hlstd{=}\hlkwd{c}\hlstd{(}\hlnum{1}\hlopt{:}\hlnum{59}\hlstd{))}\hlopt{+}
  \hlkwd{theme_bw}\hlstd{()}\hlopt{+}
  \hlkwd{theme}\hlstd{(}\hlkwc{legend.position} \hlstd{=} \hlstr{"none"}\hlstd{,}
        \hlkwc{plot.title} \hlstd{=} \hlkwd{element_text}\hlstd{(}\hlkwc{hjust} \hlstd{=} \hlnum{0.5}\hlstd{),}
        \hlkwc{strip.background} \hlstd{=} \hlkwd{element_blank}\hlstd{(),}\hlkwc{text} \hlstd{=} \hlkwd{element_text}\hlstd{(),}
        \hlkwc{axis.text.x} \hlstd{=} \hlkwd{element_blank}\hlstd{(),}
        \hlkwc{plot.margin} \hlstd{=}  \hlkwd{margin}\hlstd{(}\hlopt{-}\hlnum{0.5}\hlstd{,} \hlnum{0.5}\hlstd{,} \hlnum{0}\hlstd{,} \hlnum{0}\hlstd{,} \hlstr{"cm"}\hlstd{))}
\end{alltt}
\end{kframe}\begin{figure}
\includegraphics[width=\maxwidth]{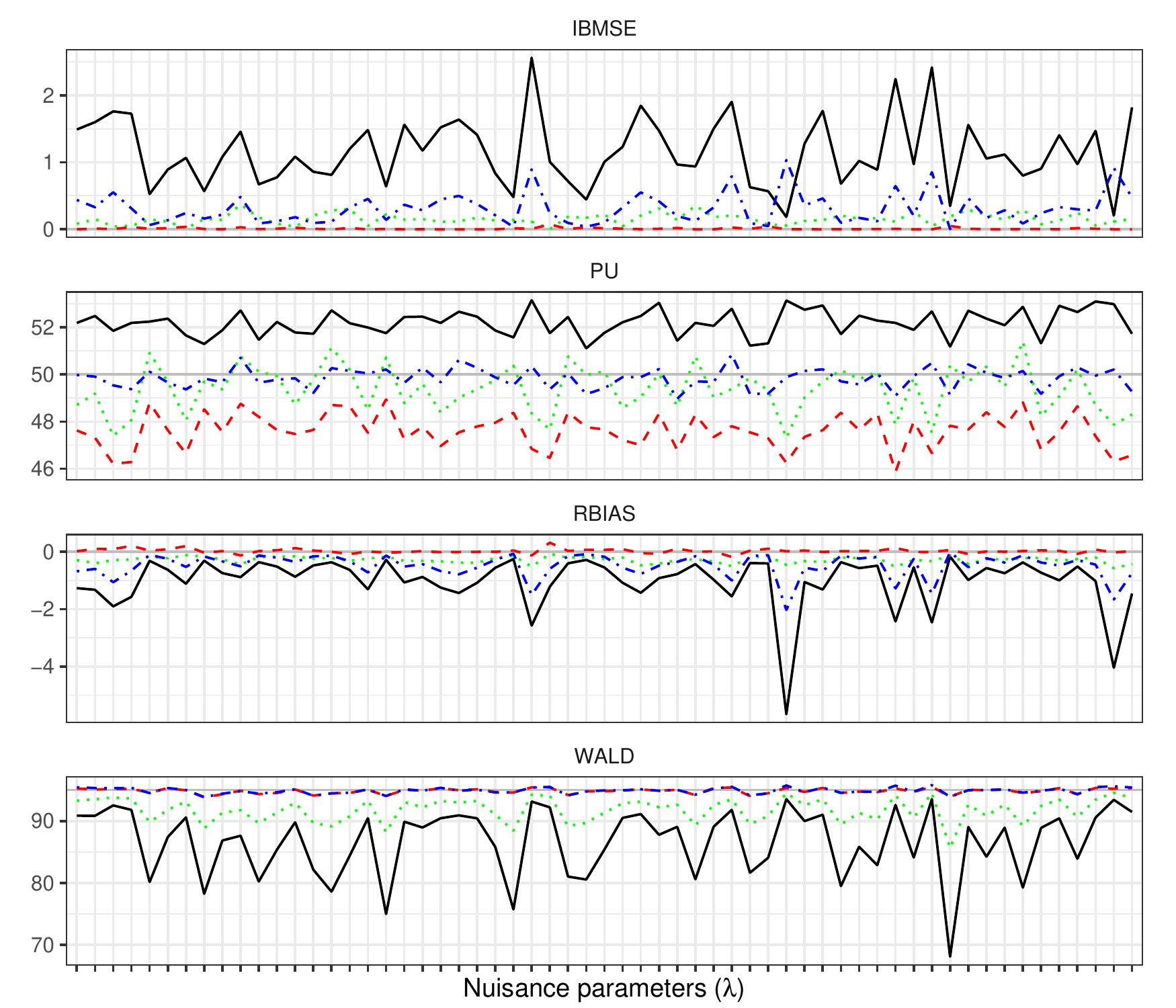} \caption[Epileptic seizures]{Epileptic seizures: Simulation results for  estimators of the nuisance parameters: ML (black solid line), mean BC (green dotted line), mean BR (red dashed line), and median BR (blue dotdash line).}\label{fig:fig8b}
\end{figure}

\end{knitrout}

\end{document}